\documentclass[manuscript]{aastex}
\usepackage{multirow}
\usepackage{graphicx,color}
\newcommand{\etal }{{et al.} }
\newcommand{\msun}{\thinspace M_\odot}

\newcommand{\vect}[1]{\mbox{\boldmath$#1$}}
\def\lesssim{\mathrel{\hbox{\rlap{\hbox{\lower4pt\hbox{$\sim$}}}\hbox{$<$}}}}
\def\gtrsim{\mathrel{\hbox{\rlap{\hbox{\lower4pt\hbox{$\sim$}}}\hbox{$>$}}}}
\newcommand{\cm}{\,{\rm cm}^{-3} } 
\newcommand{\km}{\,{\rm km\, s}^{-1}}

\newcommand{\dfrac}[2]{{\displaystyle \frac{#1}{#2}} }

\shorttitle{Different Modes of Star Formation}
\shortauthors{Machida et al.}

\begin{document}
\title{Different Modes of Star Formation: Gravitational Collapse of Magnetically Subcritical Cloud}

\author{
Masahiro N. Machida\altaffilmark{1}, Koki Higuchi\altaffilmark{1} and Satoshi Okuzumi\altaffilmark{2}
}
\altaffiltext{1}{Department of Earth and Planetary Sciences, Faculty of Sciences, Kyushu University, Fukuoka 812-8581, Japan; machida.masahiro.018@m.kyushu-u.ac.jp}
\altaffiltext{2}{Department of Earth and Planetary Sciences, Tokyo Institute of Technology, Tokyo, Japan}

\begin{abstract}
Star formation in magnetically subcritical clouds is investigated using a three-dimensional non-ideal magneto-hydrodynamics simulation.  
Since rapid cloud collapse is suppressed until the magnetic flux is sufficiently removed from the initially magnetically subcritical cloud by ambipolar diffusion, it takes $\gtrsim 5$--$10\,t_{\rm ff}$  to form a protostar, where $t_{\rm ff}$ is the freefall timescale of the initial cloud. 
The angular momentum of the star forming cloud is efficiently transferred to the interstellar medium before the rapid collapse begins, and the collapsing cloud has a very low angular momentum. 
Unlike the magnetically supercritical case, no large-scale low-velocity outflow appears in such a collapsing cloud due to the short lifetime of the first core.
Following protostar formation, a very weak high-velocity jet, which has a small momentum and might disappear at a later time, is driven near the protostar, while the circumstellar disc does not grow during the early mass accretion phase. 
The results show that the star formation process in magnetically subcritical clouds is qualitatively different from that in magnetically supercritical clouds.

\end{abstract}

\section{Introduction}
\label{sec:intro} 
Magnetic fields play an important role in the star formation process. 
In star forming regions, protostellar outflows are frequently observed
and can determine the star formation efficiency and the final stellar mass \cite[e.g.][]{nakano95}.  
The outflows are considered to be driven by magnetic effects \citep{blandford82,uchida85}. 
In a collapsing star-forming cloud, the magnetic field can transfer excess angular momentum and promote further contraction and protostar formation \citep{mestel56,shu87,basu94,tomisaka02}.

Observations indicate that although in star forming clouds the magnetic energy is comparable to the gravitational energy, it does not appear to be the case that the magnetic field supports the cloud against gravity \citep{crutcher10}. 
The mass-to-flux ratio normalized by the critical value 
\begin{equation}
\mu = \left( \dfrac{M}{\Phi} \right) / \left( \dfrac{M}{\Phi} \right)_{\rm cri},
\end{equation}
is typically used as an index of the magnetic field strength of clouds, where $\Phi$ and $M$ are the magnetic flux and mass of the cloud, respectively \citep{mouschovias76,tomisaka88a,tomisaka88b}. 
The critical value is defined as
\begin{equation}
\left( \dfrac{M}{\Phi} \right)_{\rm cri} = \dfrac{1}{2\pi G^{1/2}}.
\end{equation} 
Gravitational collapse occurs when the gravitational force overcomes the Lorentz force for $\mu>1$ (the so-called magnetically supercritical state), whereas the cloud is supported by the Lorentz force against gravity for $\mu<1$ (the so-called magnetically subcritical state). 
In the latter state, cloud contraction (or star formation) begins after the magnetic flux is removed by ambipolar diffusion \citep{mestel56,shu83,mouschovias85,basu94,basu95a,basu95b}. 
Recent observations indicate that a large fraction of clouds have $\mu>1$, while there exist clouds  with $\mu<1$ \citep{crutcher99,troland08,crutcher10}. 
The star formation and cloud contraction process in magnetically subcritical clouds was intensively investigated up until the 1990s \citep[e.g. see review by][]{shu87}.
However, in recent years, theoretical studies have mainly focused on star formation in magnetically supercritical clouds, because recent observations have suggested that the mean mass-to-flux ratio for a star forming cloud is $\mu\sim2$--$3$ \citep{troland08,crutcher10,ching17}. 
The protostar formation process in magnetically supercritical clouds has been thoroughly investigated with current state-of-the-art non-ideal magnetohydrodynamic (MHD) simulations \citep{machida07,dapp10,tomida13,tomida15,tsukamoto15,tsukamoto15b,lewis15,wurster16}. 
On the other hand, few studies of the protostar formation process in magnetically subcritical clouds has been conducted using three-dimensional non-ideal MHD simulations. 
Molecular clouds whose magnetic field strengths have been determined by observations are limited, and there are considerable uncertainties in the observations. 
By comparing the star formation process in magnetically supercritical clouds with that in magnetically subcritical clouds, we can better understand the star formation process.

In this study, we calculate the evolution of both magnetically supercritical and subcritical clouds until protostar formation to compare the star formation processes.
The paper is structured as follows. 
The initial conditions and numerical settings are described in \S\ref{sec:init} and \ref{sec:model}, respectively.
The calculation results are presented in \S\ref{sec:results}.
We discuss caveats of this study in \S\ref{sec:discussion}. 
A summary is presented in \S\ref{sec:summary}.

\section{Initial Conditions}
\label{sec:init}
The purpose of this study is to investigate the cloud collapse and star formation processes in both magnetically subcritical and supercritical clouds using a three-dimensional non-ideal MHD simulation. 
The star formation process in magnetically supercritical clouds has been investigated in many past studies, as described in \S\ref{sec:intro}. 
To directly and accurately compare cloud collapse in magnetically supercritical and subcritical clouds, we adopt almost the same initial settings as the previous studies on magnetically supercritical cases \citep{machida07,machida10a,machida11b,machida11c,machida12,machida13,machida14a,machida16}.

As the initial state (or prestellar cloud core), we adopt a critical Bonnor--Ebert sphere with a central density $ n_{\rm c} = 5 \times 10^5$\,$\cm$ and an isothermal temperature $T_{\rm iso}=10$\,K, in which a density enhancement factor $f=1.8$ is set to realize a gravitationally unstable state and induce gravitational contraction.
The initial prestellar cloud has a mass of $M_{\rm cl}=1.1\msun$ and a radius of $R_{\rm cl}= 6.2\times10^{3}$\,AU.   
The prestellar cloud is enclosed by an interstellar medium in the region of $r > R_{\rm cl}$ with a density of $n_{\rm ISM}=3.5\times10^4\cm$ and a temperature  $T_{\rm iso}=10$\,K.
The gas self-gravity is imposed only in the region of $r\le R_{\rm cl}$ to prevent gas inflow from interstellar space. 
The rigid rotation of $\Omega_0 = 1.0\times10^{-13}$\,s$^{-1}$ is set only inside the initial cloud, while the interstellar medium has no rotation. 
The ratio of rotational to gravitational energy inside the initial cloud is $\beta_0=0.01$, which corresponds to $\Omega_0\,t_{\rm ff}=0.075$, where $t_{\rm ff,0}$ ($=2.45\times10^4$\,yr) is the freefall timescale of the initial cloud centre.
A uniform magnetic field is imposed over the whole computational domain, with the magnetic field strength differing in each model (for more detailed settings, see \citealt{machida11b,machida14a,machida16}).

We prepared seven different prestellar clouds with different magnetic field strengths.
The model name, magnetic field strength and mass-to-flux ratio normalized by the critical value $(2\pi G^{1/2})^{-1}$ are listed in Table~\ref{table:1}. 
Models R01, R03 and R05 correspond to magnetically subcritical cloud models, while models R2, R3 and R5 are magnetically supercritical cloud models.  
Model R1 has a marginal mass-to-flux ratio of $\mu=1$.
Figure~\ref{fig:1} plots the initial mass-to-flux ratio for all models against the cumulative mass (upper panel) and radius (lower panel), in which the cumulative mass is radially integrated from the origin of the cloud. 
Note that the mass-to-flux ratio $\mu$ described in Table~\ref{table:1} is defined as that for the whole cloud.
As shown in the figure, the entire cloud is in a subcritical state for models R01, R03 and R05, while only the inner region of the cloud is in the subcritical state for models R1 and R2. 
On the other hand, except for a very small part of the inner cloud, almost all of the cloud is in a magnetically supercritical state for models R3 and R5.

\section{Numerical Settings}
\label{sec:model}
We calculate the evolution of both magnetically subcritical and supercritical clouds from the prestellar cloud stage up to protostar formation for the seven models listed in Table~\ref{fig:1}, using a non-ideal MHD nested grid simulation (see below). 
The basic equations are as follows:
\begin{eqnarray} 
& \dfrac{\partial \rho}{\partial t}  + \nabla \cdot (\rho \vect{v}) = 0, & \\
& \rho \dfrac{\partial \vect{v}}{\partial t} 
    + \rho(\vect{v} \cdot \nabla)\vect{v} =
    - \nabla P - \dfrac{1}{4 \pi} \vect{B} \times (\nabla \times \vect{B})
    - \rho \nabla \phi, & \\ 
\label{eq:eom}\\
& \dfrac{\partial \vect{B}}{\partial t} = 
   \nabla \times \left[  \vect{v}  \times \vect{B} - \eta_{O} (\vect{\nabla} \times \vect{B}) - \dfrac{\eta_{\rm A}}{\vert \vect{B}^2 \vert}  
\left( \vect{B} \times  (\vect{\nabla}\times \vect{B}) \right) \times \vect{B}    \right], & 
\label{eq:reg}\\
& \nabla^2 \phi = 4 \pi G \rho, &
\end{eqnarray}
where $\rho$, $\vect{v}$, $P$, $\vect{B} $, $\eta$ and $\phi$ denote the density, velocity, pressure, magnetic flux density, resistivity  and gravitational potential, respectively. 
The implementation of non-ideal terms for Ohmic resistivity and ambipolar diffusion in equation~(\ref{eq:reg}) is described in \citet{tomida15}. 
In equation (\ref{eq:reg}), the diffusion coefficients for Ohmic dissipation $\eta_{\rm O}$ and ambipolar diffusion $\eta_{\rm A}$ are taken from a pre-calculated table in the same manner as in \citet{tomida15}, in which the diffusivities depend on the density, temperature and magnetic field strength (see Appendix \S\ref{sec:A2}). 
The table is the same as that used in \citet{tomida13,tomida15} and \citet{tsukamoto15,tsukamoto15b}.
In fact, the diffusivities also depend on the assumed dust properties, though we cannot know which dust properties are more realistic and should be adopted. 
Thus, in this study, we present qualitative results, because physical quantities such as the magnetic flux and angular momentum in a collapsing cloud might depend on the assumed dust properties.

To mimic the temperature evolution \citep{larson69,masunaga00}, we adopt the piece-wise polytropic equation of state \citep[see][]{tomisaka02,machida07,machida11c,dapp12,lewis15,wurster16} as 
\begin{equation} 
P = \left\{
\begin{array}{ll}
 c_{s,0}^2 \rho & \rho < \rho_c, \\
 c_{s,0}^2 \rho_c \left( \frac{\rho}{\rho_c}\right)^{7/5} &\rho_c < \rho < \rho_d, \\
 c_{s,0}^2 \rho_c \left( \frac{\rho_d}{\rho_c}\right)^{7/5} \left( \frac{\rho}{\rho_d} \right)^{1.1}
 & \rho_d < \rho < \rho_e, \\
 c_{s,0}^2 \rho_c \left( \frac{\rho_d}{\rho_c}\right)^{7/5} \left( \frac{\rho_e}{\rho_d} \right)^{1.1}
 \left( \frac{\rho}{\rho_e}   \right)^{5/3}
 & \rho > \rho_e, 
\label{eq:eos}
\end{array}
\right.  
\end{equation}
 where $c_{s,0} = 190$\,m\,s$^{-1}$, 
$ \rho_c = 3.84 \times 10^{-13} \, \rm{g} \, \cm$ ($n_c \simeq 10^{11} \cm$), 
$ \rho_d = 3.84 \times 10^{-8} \, \rm{g} \, \cm$  ($n_d \simeq  10^{16} \cm$) and
$ \rho_e = 3.84 \times 10^{-3} \, \rm{g} \, \cm$  ($n_e \simeq 10^{21} \cm$).
Because the barotropic equation of state is used, we ignore heating by Ohmic dissipation and ambipolar diffusion in this study (for details, see Appendix \S\ref{sec:appendix}).

Since a detailed description of the nested grid code has been given in our previous papers \citep{machida04,machida05a,machida05b,machida07,machida10a,machida13}, we briefly summarize the calculation code.
Before starting the calculation, we prepared six levels of nested rectangular grids ($l=1$--$6$ ), in which the grid size $L(l)$ and cell width $h(l)$ halve with each increment of the grid level $l$.  
Each grid is composed of (i, j, k) = (64, 64, 32)  and a mirror symmetry is imposed on the $z=0$ plane. 
The grid size and cell width of the $l=1$ grid are $L(1)=2\times 10^5\,$AU and $h(1)=3100$\,AU, respectively.
The initial cloud is enclosed in the $l=5$ grid level, and the interstellar medium is set outside the initial cloud. 
Thus, we set a wide region of interstellar space (16 times larger than the prestellar cloud radius) to prevent the reflection of the Alfv\'en wave from the computational boundary \citep{machida06}. 
A new grid level is dynamically generated before the Truelove condition is violated \citep{truelove97}, in which the Jeans length is resolved in at least 16 cells . 
The maximum grid level is set to $l=21$ which has $L(21)=0.19$\,AU and $h(21)=2.9\times10^{-3}$\,AU. 
With these settings, we calculate the cloud evolution until the central density reaches $n_{\rm c}\sim10^{22}\cm$.

\section{Results} 
\label{sec:results}
\subsection{Removal of magnetic field and quasi-static state}
\label{sec:results1}
Figure~\ref{fig:2} shows the time evolution of the central density for each model. 
The central density rapidly increases and a protostar forms in $t \sim2$--$3\,  t_{\rm ff,0}$ for the magnetically supercritical models R2, R3 and R5. 
On the other hand, the magnetically subcritical models R01, R03 and R05 require  $t \sim 5$--$10\, t_{\rm ff,0}$ in order to form a protostar, because cloud collapse is not induced until the magnetic flux is sufficiently removed from the central region by ambipolar diffusion. 
The figure clearly indicates that a strong magnetic field delays star formation, as also shown in recent studies \citep{tassis07, kunz10}.

\begin{figure*}
\includegraphics[width=150mm]{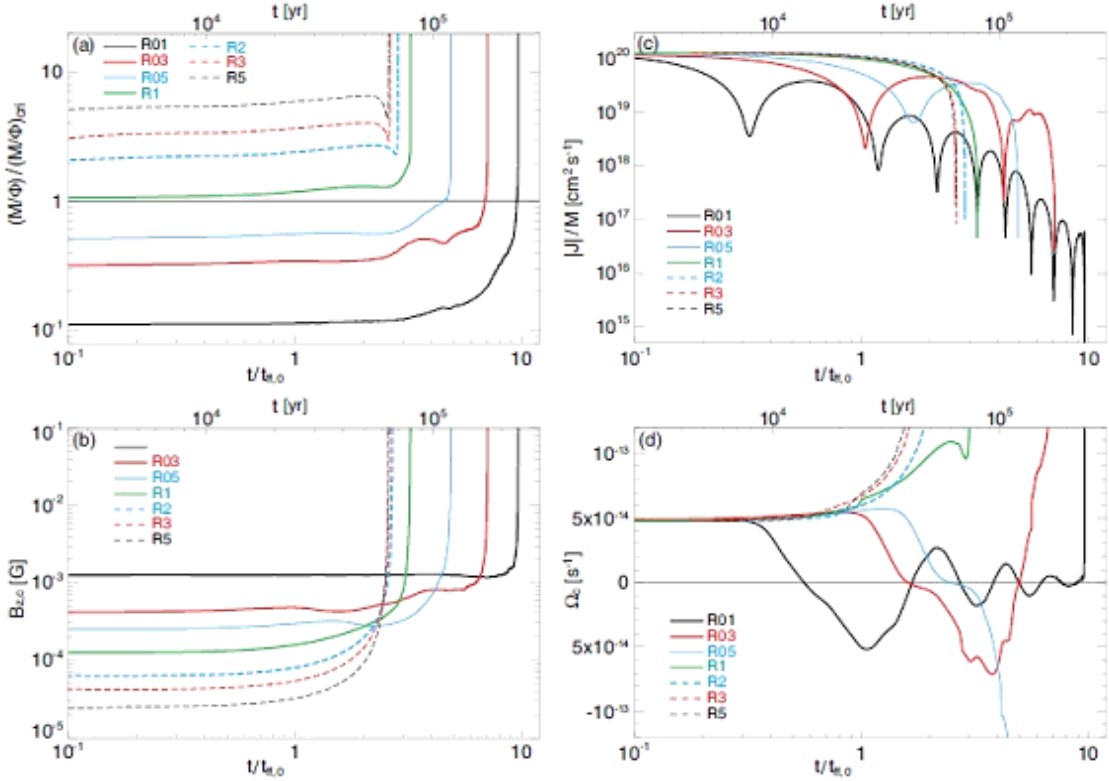}
\caption{
Time evolution of (a) mass-to-flux ratio in the region $\rho > 0.1\rho_{\rm max}$, (b) magnetic field strength at the centre of the cloud, 
(c) absolute value of the specific angular momentum in the region $\rho > 0.1\rho_{\rm max}$ and (d) angular velocity at the centre of the cloud.   
In each panel, the $x$-axis corresponds to the elapsed time normalized by the initial free fall timescale (lower axis) and that in unit of years (upper axis).
}
\label{fig:3}
\end{figure*}

The physical quantities are plotted against the elapsed time normalized by the initial freefall timescale in Figure~\ref{fig:3}.
Figure~\ref{fig:3}{\it a} shows the mass-to-flux ratio normalized by the critical value estimated in the region $\rho>0.1\, \rho_{\rm max}$ and indicates that rapid cloud collapse begins when the normalized mass-to-flux ratio exceeds $\mu \gtrsim 1$ for the magnetically subcritical models R01, R03 and R05, where $\mu$ is defined as $(M/\Phi)/(M/\Phi)_{\rm cri}$ in the region of $\rho > 0.1\, \rho_{\rm max}$. 
In these models, neutral gases slip through magnetic field lines by ambipolar diffusion.
Thus, rapid collapse is induced after the magnetic flux is removed from the central region and gravity overcomes the Lorentz force.
Figure~\ref{fig:3}{\it b} shows the evolution of the magnetic field strength at the cloud centre. 
Before the rapid collapse occurs, each model has an almost constant value of magnetic field strength. 
A uniform magnetic field is adopted as the initial state. 
With an ambipolar diffusion term (eq.~[\ref{eq:reg}]), a uniform magnetic field distribution tends to be realized.
Thus, the magnetic field lines do not significantly move, while neutral gases slip through the magnetic field lines and are gradually concentrated toward the cloud centre.
Therefore, the mass-to-flux ratio decreases and rapid collapse occurs even for magnetically subcritical models.

Figures~\ref{fig:3} right panels show the specific angular momentum in the region $\rho>0.1\, \rho_{\rm max}$ (Fig.~\ref{fig:3}{\it c}) and angular velocity at the cloud centre (Fig.~\ref{fig:3}{\it d}). 
The angular velocities gradually increase as the collapse proceeds in magnetically supercritical clouds because the angular momentum is (roughly) conserved (Fig.~\ref{fig:3}{\it d}).
On the other hand, in magnetically subcritical clouds, the angular momentum of the cloud  is transferred to the interstellar space by magnetic braking before the rapid collapse begins \citep{nakano90,basu94,basu95a}. 
Figure~\ref{fig:3}{\it d} shows that for magnetically subcritical models, the angular velocity decreases in an oscillating fashion before the rapid collapse occurs. 
The oscillation of the angular velocity is due to inversion of the angular momentum vector (angular momentum vector parallel or anti-parallel to the $z$-axis). 
The neutral gases are partially coupled with the magnetic field lines which penetrate the cloud  and are anchored to the interstellar medium. 
Thus, after the magnetic field lines are twisted by the initial rotation motion of the cloud, they are swung back by the magnetic tension force. 
Then, the magnetic field lines are oppositely twisted toward the initial rotation direction, and neutral gases, which are partly coupled with the magnetic field lines, have an anti-rotation motion toward the initial rotation motion.
Therefore, although the absolute value of the angular momentum decreases with time due to the magnetic braking, the swing back of the magnetic field lines and inversion of the rotation vector are repeated before the rapid collapse occurs (or until the magnetic fluxes are sufficiently removed from the cloud).
The decrease and oscillation of the angular velocity have already been presented in classical works \citep{mouschovias79,mouschovias80}.

Due to the turnover of the rotation vector when the rapid collapse begins, the direction of rotation motion at the cloud centre is randomly determined in magnetically subcritical models. 
As seen in Figure~\ref{fig:3}{\it d}, for magnetically subcritical models, the rotation vector at the cloud centre is parallel to the $z$-axis (e.g. $\Omega_{\rm c}>0$) for models R01 and R03, while model R05 has an anti-parallel rotation (e.g. $\Omega_{\rm c}<0$). 
The direction of the rotation vector (parallel or anti-parallel to the initial rotation axis) is determined by the time at which the rapid cloud collapse begins when $\mu\gtrsim 1$.

\begin{figure*}
\includegraphics[width=150mm]{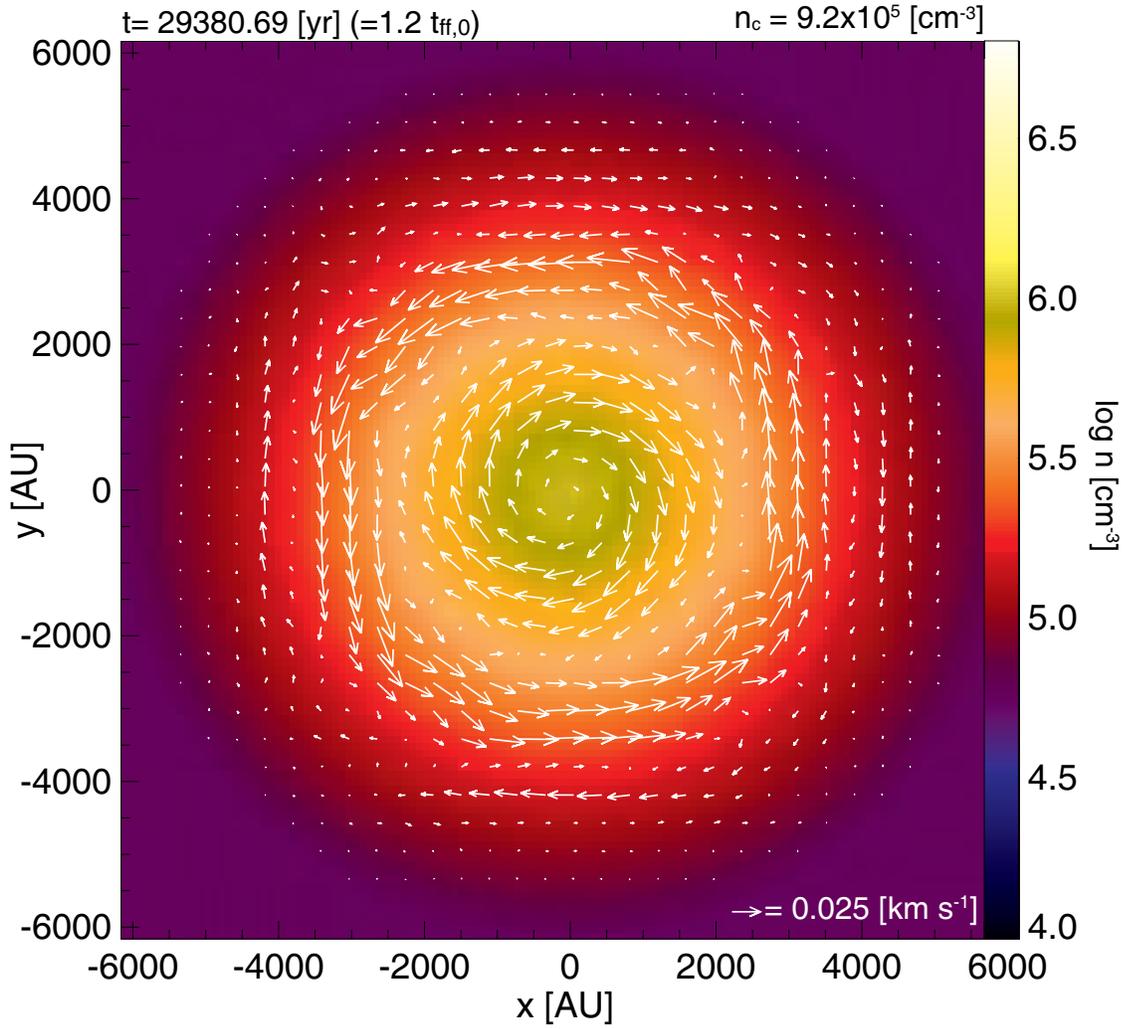}
\caption{
Density (colour) and velocity (arrows) distributions on the equatorial plane before rapid collapse occurs for model R01.
}
\label{fig:4}
\end{figure*}

Although only the central part of the rotation motion is shown in Figure~\ref{fig:3} right panels, the rotation motion differs at each radius in a magnetically subcritical cloud, as shown in Figure~\ref{fig:4}. 
In the figure, clockwise and anti-clockwise rotation coexist in the star forming cloud because the momentum inertia at each radius differs.
Note that the efficiency of the magnetic braking is determined by the momentum inertia in the region where the Alfv\'en wave reaches in a time  \citep{mouschovias79,basu94}. 
Thus, it is expected that lumps of gas with different directions of angular vector fall onto the central region after protostar formation.

\subsection{Cloud collapse before protostar formation}
\label{sec:beforeps}
\begin{figure*}
\includegraphics[width=140mm]{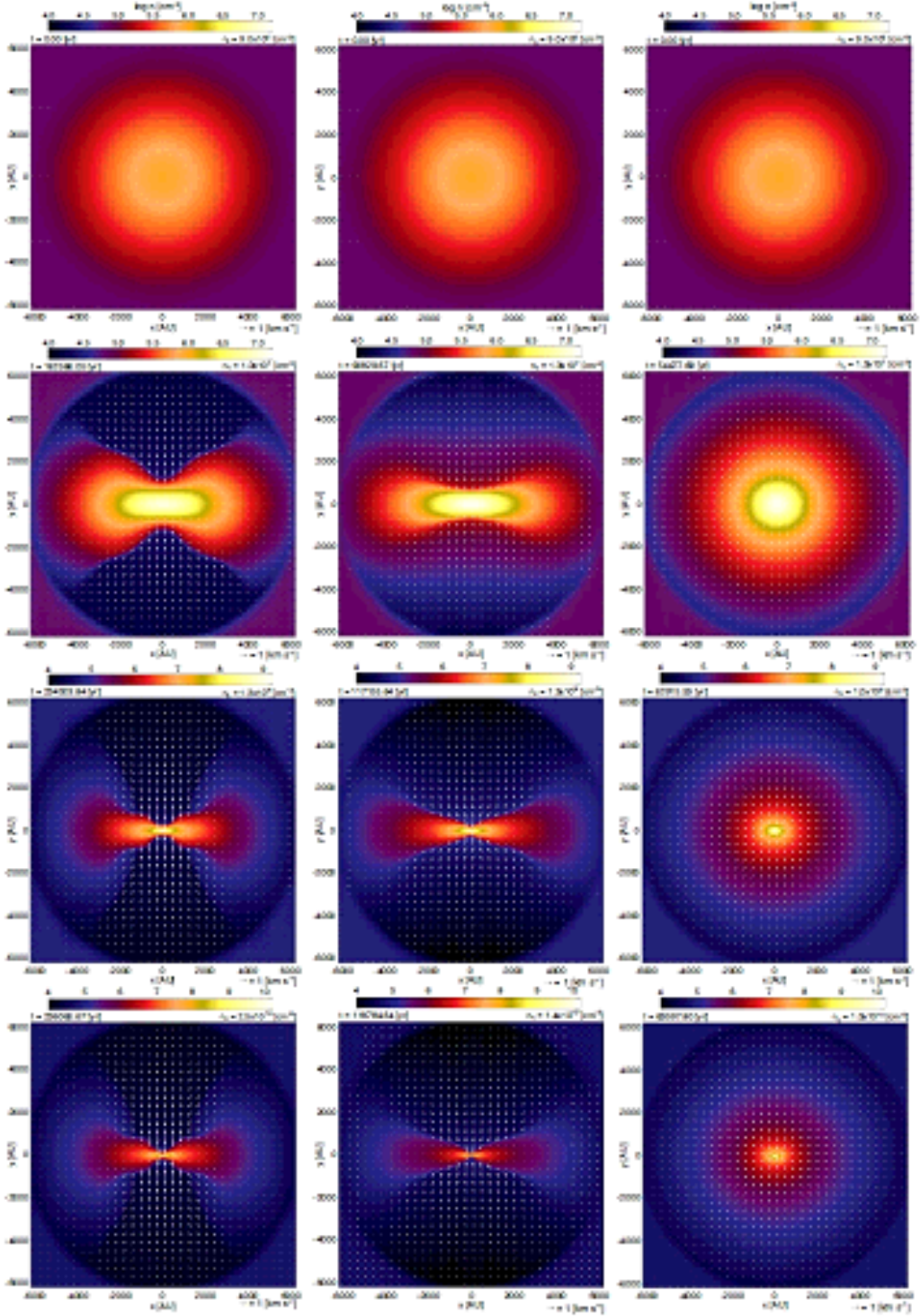}
\caption{
Time sequence of models R01 (left),  R05 (middle) and R3 (right). 
The density (colour) and velocity (arrows) distributions on the $x=0$ plane are plotted in each panel.
The elapsed time $t$ and central number density $n_{\rm c}$ are indicated above each panel. 
}
\label{fig:5}
\end{figure*}

Figure~\ref{fig:5} shows the time sequence of cloud shapes for models R01 (left column), R05 (middle column) and R3 (right column), in which the central density is almost the same at each row. 
The cloud collapses spherically at a large scale for the magnetically supercritical model R3 (right column). 
On the other hand, the whole cloud is transformed from spherical to a disc-like (or butterfly-like) structure for subcritical models R01 and R05, in which the cloud contracts quasi-statically. 
The cloud structure for magnetically subcritical models is almost the same as that seen in past studies \citep{nakano82,tomisaka88b,tomisaka89,tomisaka90}. 
Thus, the large scale cloud configuration differs considerably between magnetically subcritical and supercritical clouds. 
The difference in cloud configuration is expected to influence the formation and evolution of circumstellar discs, as shown in \citet{machida16}.

Various physical quantities for each model in the gas collapsing phase are plotted against the central number density in Figure~\ref{fig:6}. 
Figure~\ref{fig:6} left panels indicate that the angular momentum and angular velocity in magnetically subcritical clouds are considerably smaller than those in magnetically supercritical clouds. 
As described in \S\ref{sec:results1}, the angular momentum is significantly removed from the cloud  before rapid collapse occurs. 
In addition, in magnetically subcritical clouds, a strong magnetic field continues to transfer the angular momentum outward even in the gas collapsing phase \citep[e.g.][]{basu94,basu95a}. 
Thus, even when the initial clouds have the same angular momentum, the angular momentum in magnetically subcritical clouds is significantly smaller than that in magnetically supercritical clouds.

Figure~\ref{fig:6} right panels indicate that, in the gas collapsing phase,  the magnetic field in magnetically subcritical clouds is more amplified than that in magnetically supercritical clouds. 
Thus, clouds with an initially strong magnetic field produce protostars having strong magnetic fields. 
The amplification of the magnetic field in a collapsing cloud is closely related to the first core lifetime and the magnetic dissipation inside it (see, \S\ref{sec:appendix}).

In the gas collapse phase, the first core forms \citep{larson69,masunaga00}. 
The first core is in a quasi-hydrostatic state and the central density gradually increases as its mass increases by mass accretion. 
After the central density exceeds $n_{\rm c} \gtrsim 10^{16}\cm$, the molecular hydrogen begins to dissociate and a second collapse occurs. 
The lifetime of the first core is about $10^2$--$10^4$\,yr and strongly depends on the rotation rate of the first core \citep{saigo06,saigo08,tomida10}. 
Note that the first core lifetime is roughly determined by $ t_{\rm fc}= M_{\rm fc} /(\dot{M}_{\rm fc})$, where $M_{\rm fc}$ and $\dot{M}_{\rm fc}$ are the mass of the first core and the mass accretion rate onto the first core, and the first core becomes massive with rotation \citep{saigo06}.
Thus, assuming a constant mass accretion rate, the rotating first core has a longer lifetime than the non-rotating first core.
In magnetically supercritical clouds, the magnetic field efficiently dissipates both by Ohmic dissipation and ambipolar diffusion inside (or around) the first core, because the ionisation degree is extremely low \citep[e.g.][]{machida07,tomida13,tomida15,tsukamoto15b}. 
Therefore, magnetic braking is alleviated in the first core and the first core remnant evolves into a rotationally supported disc  following protostar formation \citep{bate98,bate10,walch09,machida11c,walch12}. 
In addition, the low-velocity outflow is driven by the outer edge of the rotating first core where the magnetic field is coupled with neutral gas \citep{tomisaka02,banerjee06,machida08,duffin09}.

On the other hand, the lifetime of the first core is as short as $t \lesssim 100$\,yr when the angular momentum of the first core is sufficiently small \citep{masunaga00,saigo06,saigo08}. 
In such a case, there is not enough time for dissipation of the magnetic field, and the magnetic field continues to be amplified without dissipation \citep{machida07}. 
As a result, the amplification rate for the magnetic field depends on the first core lifetime or rotation rate.
Thus, as shown in Figure~\ref{fig:6}{\it f}, the magnetic fields in magnetically subcritical models are stronger than those in magnetically supercritical models when the central density reaches $n_{\rm c}\sim10^{12}-10^{14}\cm$, at which point the first core forms and dissipation of the magnetic field again becomes effective \citep{nakano02}. 
Note that although the magnetic field strengths are normalized by $(8\pi c_{s,0}^2 \rho_{\rm c})$ to stress their difference in Figure~\ref{fig:6}{\it f}, the difference in the magnetic fields $B_{z,0}$ among models is not very large, as seen in Figure~\ref{fig:6}{\it e}.

\begin{figure*}
\includegraphics[width=150mm]{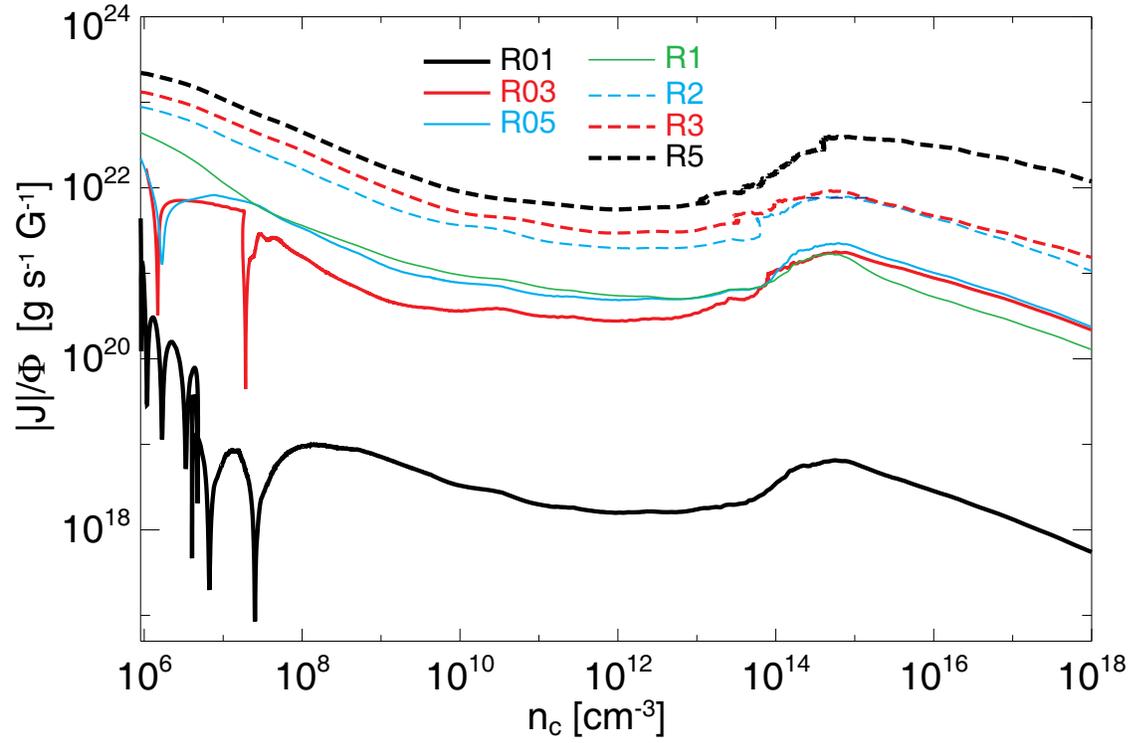}
\caption{
Ratio of angular momentum to magnetic flux in the region $\rho > 0.1 \rho_{\rm max}$ against the central number density.
}
\label{fig:7}
\end{figure*}

To illustrate the relation between the rotation and magnetic field, the ratio of the angular momentum to the magnetic flux for each model is plotted against the central number density in Figure~\ref{fig:7}.
The figure clearly indicates that the magnetically supercritical clouds have a large angular momentum but a relatively weak magnetic field. 
On the other hand, a strong magnetic field and small angular momentum are realized in magnetically subcritical clouds. 
The difference of $\vert J \vert / \Phi$ causes dichotomization of the star formation mode (\S\ref{sec:summary}).

\subsection{Protostar formation epoch}
\begin{figure*}
\includegraphics[width=150mm]{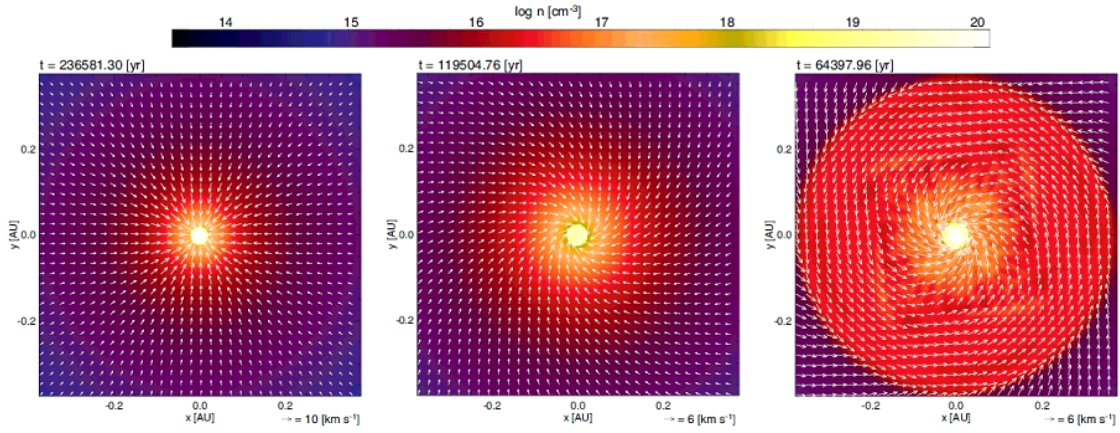}
\caption{
Density (colour) and velocity (arrows) distributions on the equatorial plane at the protostar formation epoch for models R01 (left), R05 (middle) and R3 (right).
The elapsed time after the cloud collapse begins is indicated above each panel. 
}
\label{fig:8}
\end{figure*}

Figure~\ref{fig:8} shows the density and velocity distributions at the protostar formation epoch for models R01, R05 and R3. 
In each panel, the central white region corresponds to a protostar with a size of $\sim0.01$\,AU and a mass of $\sim0.01\msun$ when the central density reaches $n_{\rm c}\sim 10^{20}\cm$.
The gas radially falls onto the protostar without rotation in model R01, which has the strongest initial magnetic field, and almost all the initial angular momentum has already been transferred into the interstellar medium before the rapid collapse begins (\S\ref{sec:results1}). 
For model R01, no clear circumstellar disc appears.  
For model R05 (middle panel), although rotational motion is confirmed, the infalling material is not supported by the centrifugal force. 
Thus, the gas spirals into the protostar.
In the panel, the rotation direction (clockwise rotation) is opposite to the initial rotation direction (anti-clockwise rotation).

A rotationally supported disc is confirmed for the magnetically supercritical model R3 (right panel). 
As shown in previous studies \citep{machida11c,machida16}, the first core remnant evolves into a circumstellar disc in a magnetically supercritical cloud. 
This is because the first core survives for a long time with rotation and the magnetic field in the first core dissipates. 
Therefore, the magnetic braking weakens and the first core is mainly supported by the centrifugal force \citep{tomida13,tomida15,tsukamoto15b}. 
Then, after protostar formation, the first core evolves into a rotationally supported circumstellar disc. 
In the right panel of Figure~\ref{fig:8}, although the size of the circumstellar disc is as small as $\sim0.5$\,AU, we can confirm that the rotational motion dominates the infall motion in the range  $r\sim0.2$--$0.5$\,AU. 
Figure~\ref{fig:8} indicates that the circumstances around protostars differ considerably between magnetically subcritical and supercritical models.

\subsection{Gas accretion phase just after protostar formation}
\begin{figure*}
\includegraphics[width=150mm]{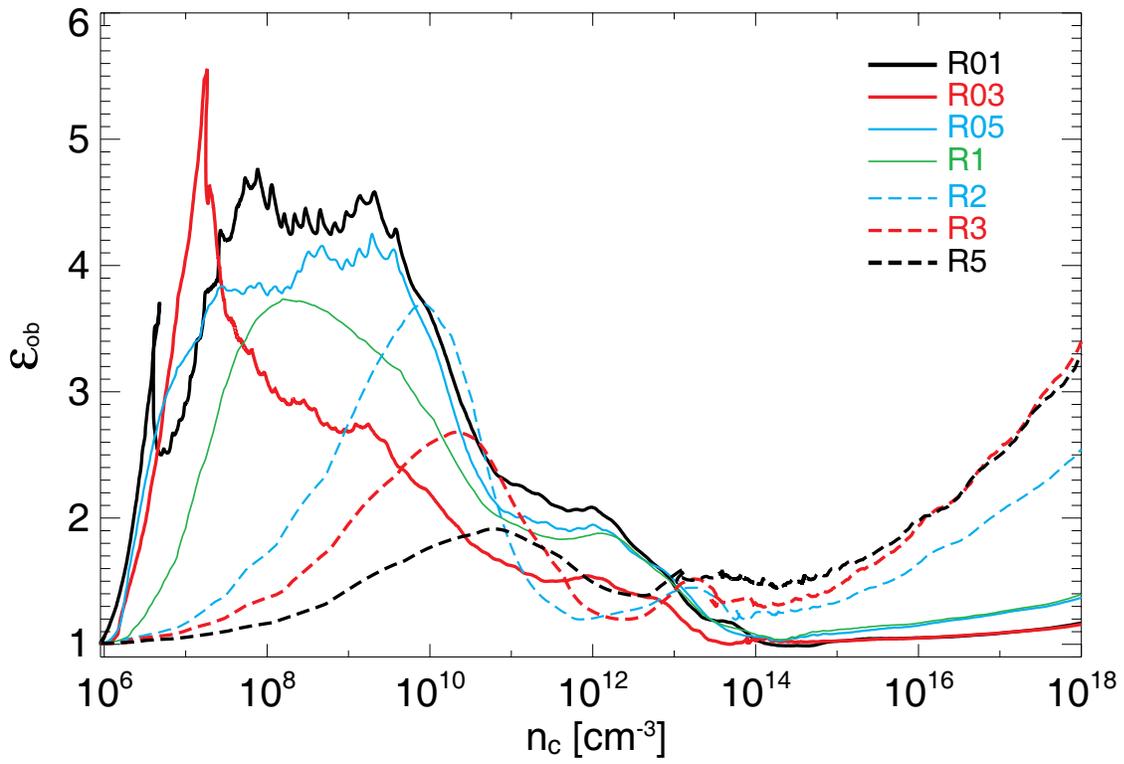}
\caption{
Oblateness for each model plotted against the central number density.
}
\label{fig:9}
\end{figure*}

Figure~\ref{fig:9} shows the oblateness for each model plotted against the central number density.
The oblateness is used as an index of the degree of flatness of an object. A larger oblateness indicates a more flattened structure. 
The oblateness is estimated as $\varepsilon_{\rm ob}=(h_l\,h_s)^{1/2}/h_z$, where $h_l$ and $h_s$ are the long and short axis and $h_z$ is the $z$-axis in the region $\rho>0.1 \rho_{\rm max}$. 
The axes $h_l$, $h_s$ and $h_z$ are calculated using the moment of inertia (for details, see \citealt{matsu99,machida05a,machida16}). 
Figure~\ref{fig:9} indicates that, for magnetically subcritical models, the cloud has a flattened structure at large scales (or lower density of $n_{\rm c} \lesssim 10^{10}\cm$) and a spherical shape around the protostar (or higher density of $n_{\rm c} \gtrsim 10^{14}\cm$). 
For these models, a strong magnetic field produces a flattened disc at large scales, while at small scales, the anisotropic Lorentz and centrifugal forces are weak due to magnetic dissipation and removal of the angular velocity, and a spherical structure forms. 
For magnetically supercritical models, a flattened structure, corresponding to a pseudo-disc created by the Lorentz force, appears in the range $10^8\lesssim n_{\rm c} \lesssim 10^{11}\cm$. 
Then, although the oblateness decreases immediately after first core formation (or the scale of the first core), a dense region ($n\gtrsim 10^{14}\cm$) transforms into a disc-like structure, which corresponds to a rotationally supported disc. 
As a result, the cloud structure differs considerably between magnetically subcritical and supercritical models.

\begin{figure*}
\includegraphics[width=150mm]{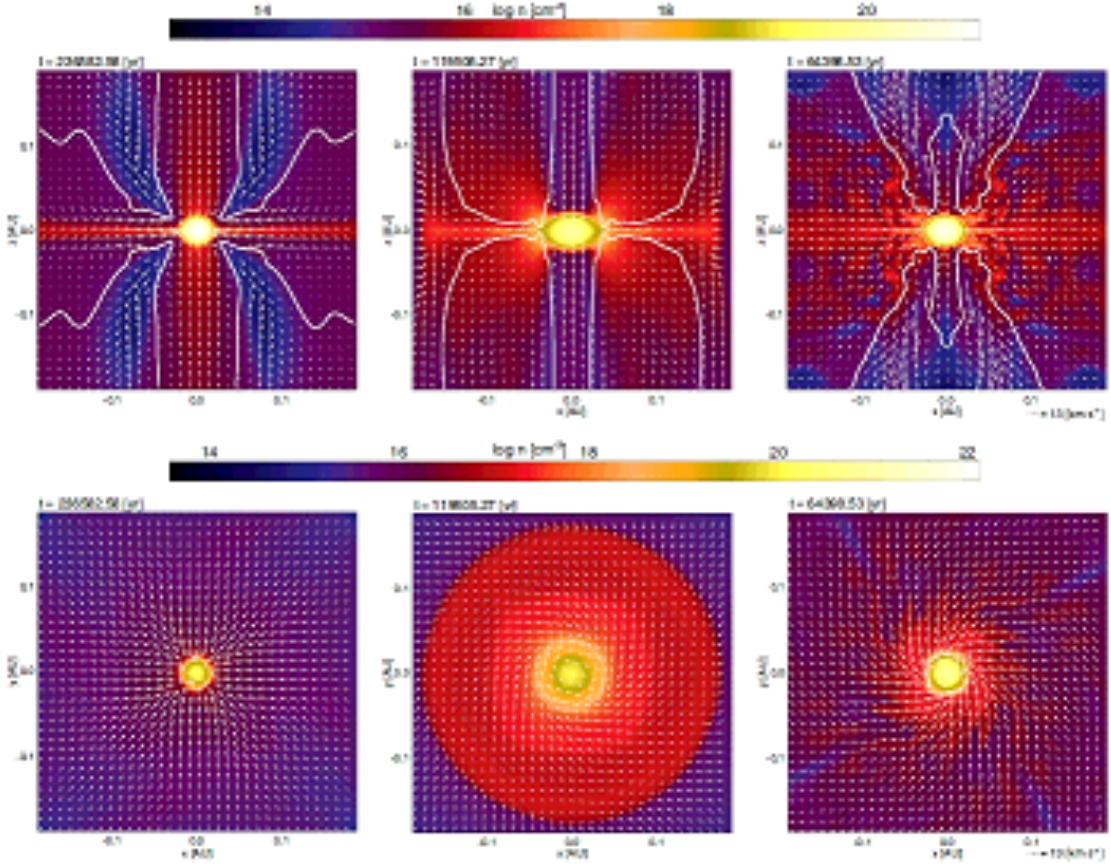}
\caption{
Density (colour) and velocity (arrows) distributions on the $y=0$ plane (top panels) and on the equatorial plane (bottom panels) during the gas accretion stage for models R01 (left), R05 (middle) and R3 (right). 
The white contours in the top panels indicate the boundary between the infalling and outflowing gases. 
The elapsed time after the cloud collapse begins is indicated above each panel. 
}
\label{fig:10}
\end{figure*}

Figure~\ref{fig:10} shows the density and velocity distributions around a protostar in the gas accretion stage. 
In model R01, the gas collapse stops on the protostar surface and the radial velocity becomes nearly zero. 
Thus, even when the rotation velocity is very small, the rotation generates a toroidal field and the magnetic fields are amplified around the protostar. 
As a result, mass ejection occurs by magnetic effects, as seen in magnetically supercritical models (right panel). 
For model R05 (middle panel), during the gas accretion phase, since lumps of gas with a relatively high angular momentum (see Figs.~\ref{fig:3} and \ref{fig:6}) fall near the protostar, the rotationally supported disc seems to gradually grow. 
In addition, mass ejection occurs near the surface of the circumstellar disc. 

For the magnetically supercritical model R3 (right panel), strong mass ejection occurs near the protostar, as seen in previous works \citep{tomisaka02,banerjee06,machida08}. 
In this epoch, although the rotationally supported disc extends up to $\sim0.5$\,AU, the angular momentum around the protostar is effectively transferred by magnetic effects \citep{machida14b}. 
Thus, the gravitational energy of the disc is released and strong mass ejection occurs.

\begin{figure*}
\includegraphics[width=150mm]{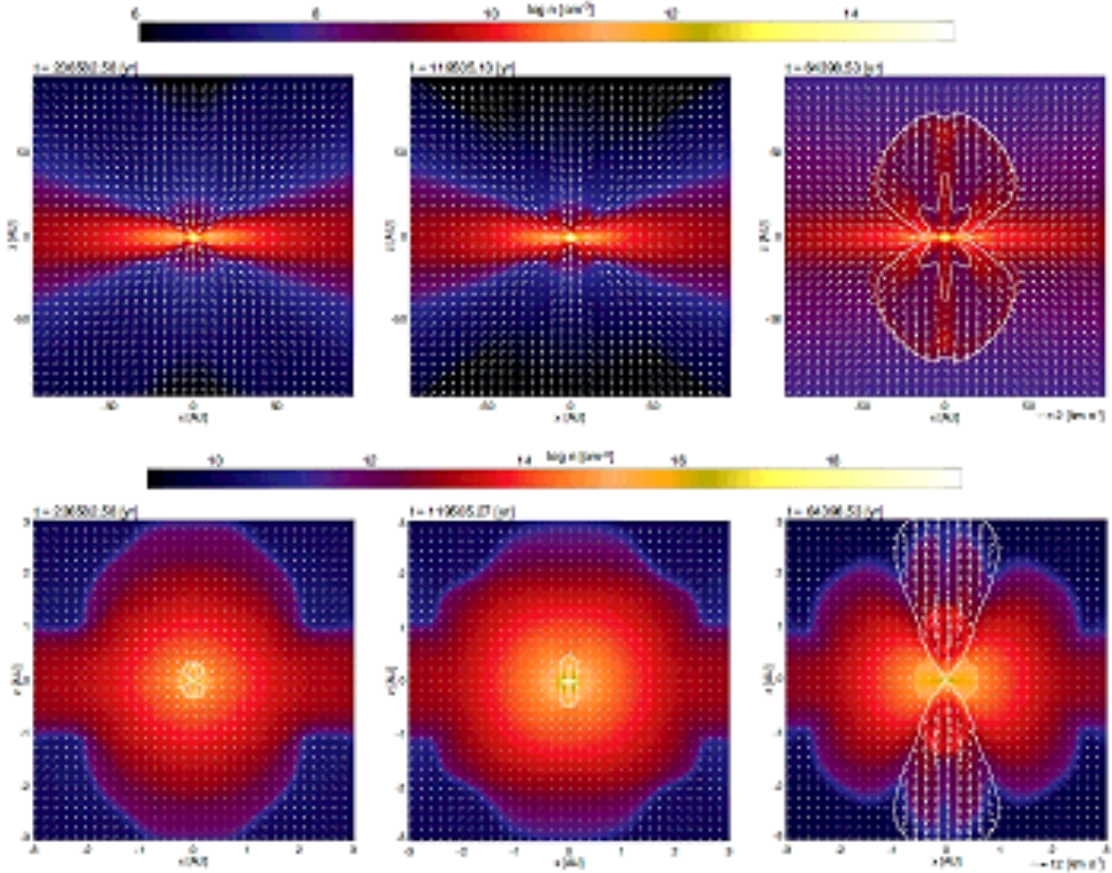}
\caption{
Density (colour) and velocity (arrows) distributions on the $y=0$ plane during the gas accretion stage for models R01 (left), R05 (middle) and R3 (right). 
The box scales for the top and bottom panels are $200$\,AU and $6$\,AU, respectively. 
The white contours in each panel indicate the boundary between the infalling and outflowing gases. 
The elapsed time after the cloud collapse begins is indicated above each panel. 
}
\label{fig:11}
\end{figure*}

The top panels in Figure~\ref{fig:11} show a large scale structure during the mass accretion stage for models R01 (left), R05 (middle) and R3 (right). 
In the magnetically subcritical models R01 and R03, although a disc-like structure is seen, no large scale outflow appears.
On the other hand, a largescale outflow reaches $\sim70$\,AU for the magnetically supercritical model. 
As seen in many previous works, a low-velocity outflow is driven by the first core that evolves into a circumstellar disc after protostar formation \citep[e.g.][]{tomida15}.
Thus, an outflow appears and extends out a large distance before protostar formation. 
At small scales (Fig.~\ref{fig:11} lower panels), strong mass ejection can be seen in the magnetically supercritical model, while weak flows appear in the magnetically subcritical models.

\begin{figure*}
\includegraphics[width=150mm]{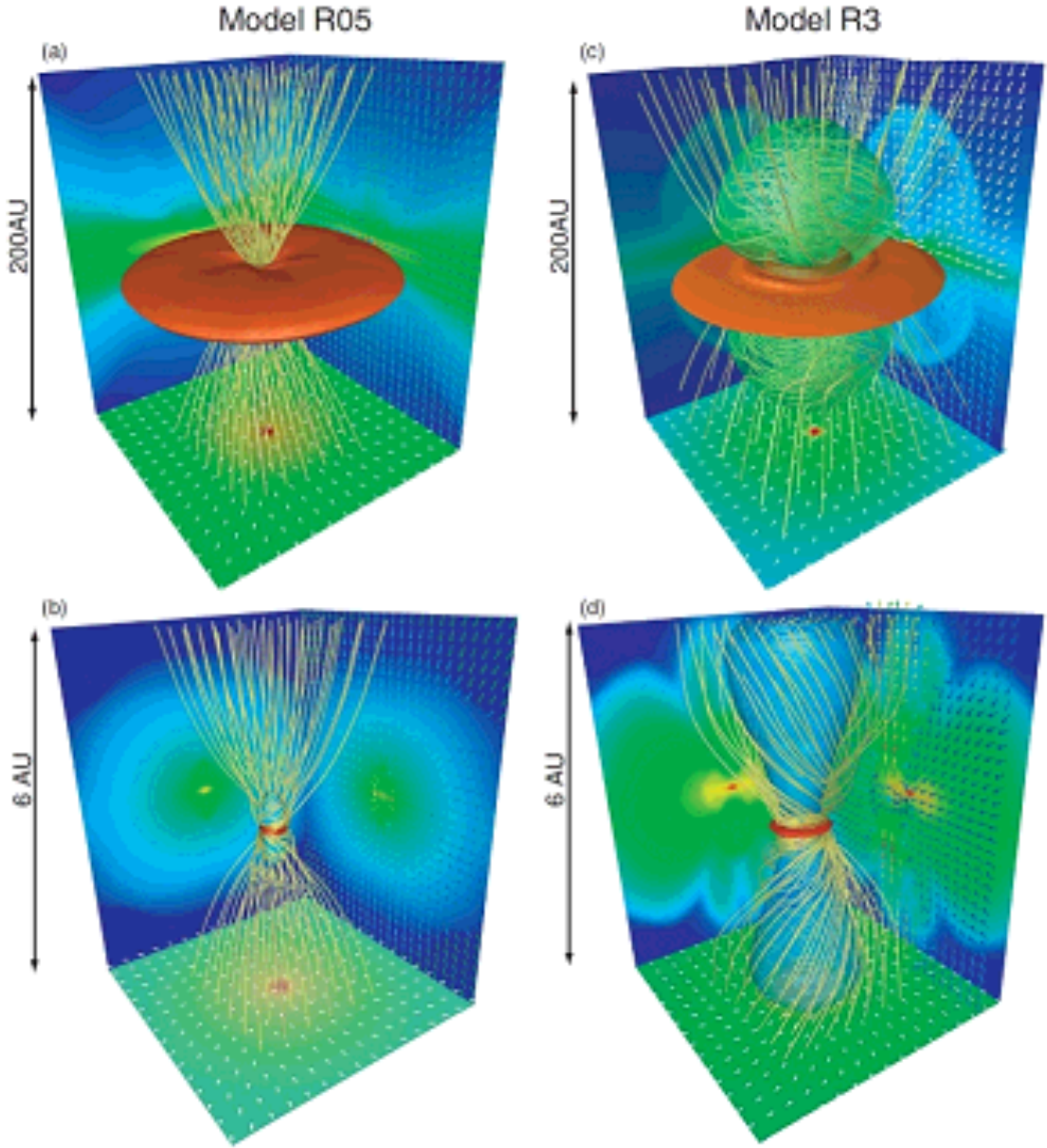}
\caption{
Three-dimensional view of large (top panels) and small (bottom panels) scale structures of models R05 (left panels) and R3 (right panels).
The yellow lines are the magnetic field lines. 
The red region corresponds to the high-density region.
The outflowing region is indicated by the green (top panel) and blue (bottom panel) surfaces. 
The spatial scale is plotted in each panel.
The density distribution and velocity vectors are projected on each wall. 
}
\label{fig:12}
\end{figure*}

Figure~\ref{fig:12} shows three-dimensional views of models R05 (left panels) and R3 (right panels). 
For model R05, the magnetic field lines converge toward the centre and are not significantly twisted (Fig.~\ref{fig:12}{\it a}).
At small scales (Fig.~\ref{fig:12}{\it b}), the magnetic field lines are twisted around the protostar and a weak jet can been seen. 
For the magnetically supercritical model R3, the magnetic field lines are strongly twisted and mass ejection occurs at both large and small scales (Fig.~\ref{fig:12}{\it c} and {\it d}).

Figure~\ref{fig:13} shows the time evolution of outflow and jet momenta, calculated by
\begin{equation}
P_{\rm Out} = \int_{0.2< v_r/(\km) \le 2} \rho \,  \vect{v} \, dV,
\end{equation}
\begin{equation}
P_{\rm Jet} = \int_{v_r/(\km) > 2} \rho \, \vect{v} \, dV,
\end{equation}
in which the low ($P_{\rm Out}$) and high ($P_{\rm Jet}$) velocity components are estimated (for details, see \citealt{machida14b}).  
The protostar formation epoch $t_{\rm ps}$ is defined as when the central density reaches $n_{\rm c}=10^{20}\cm$.
The figure indicates that for the magnetically supercritical models R2 and R3, low-velocity outflow appears before protostar formation and continues to be driven after protostar formation.
On the other hand, no low-velocity component appears before protostar formation for the magnetically subcritical models R03 and R05. 
Note that the low-velocity component ($0.2\km < v_r < 2\km$) appears after protostar formation even for models R03 and R05, but the outflow momentum is significantly lower than $P_{\rm out}<10^{-5}\msun\km$. 
Note also that, in models R03 and R05, a low-velocity component is entrained by the high-velocity component.

\subsection{Parameter dependence and further evolution}
\begin{figure*}
\includegraphics[width=150mm]{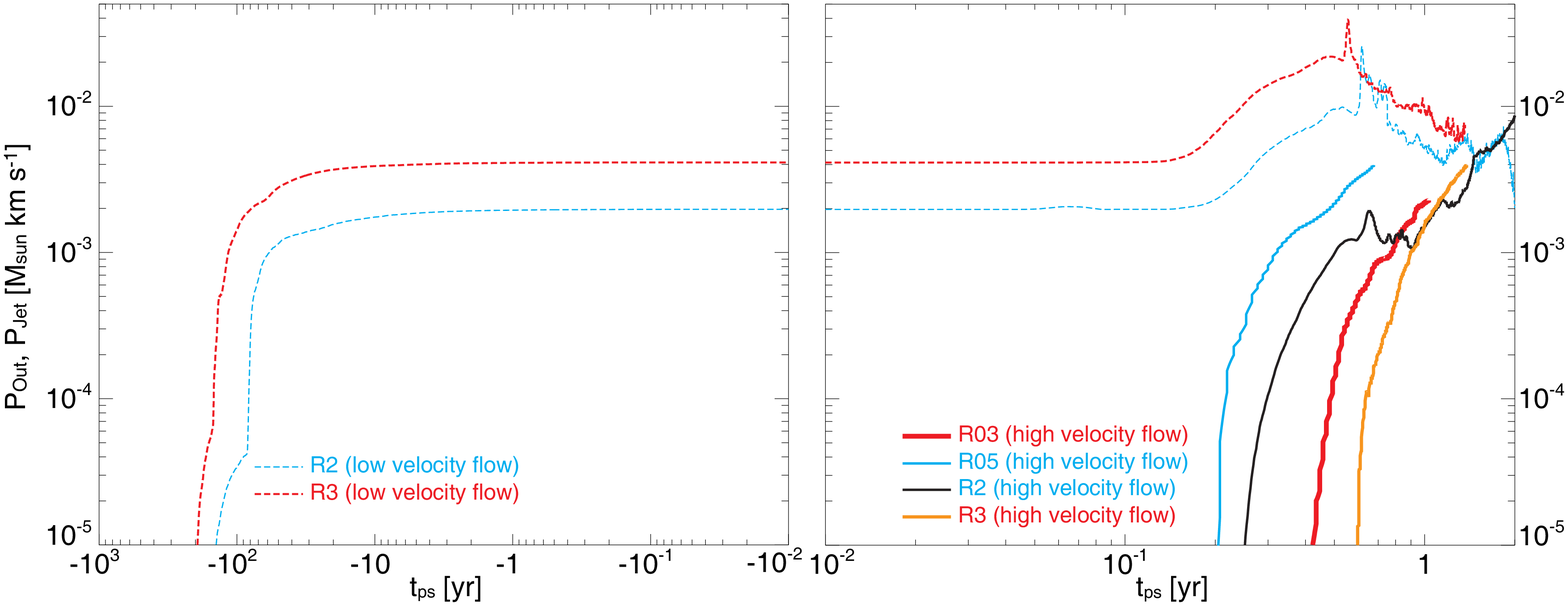}
\caption{
Momentum of low (broken lines: outflow) and high (solid lines: jet) velocity outflows for models R03, R05, R2 and R3 against the elapsed time. 
The origin of time ($t_{\rm ps}$) is set to the protostar formation epoch. 
}
\label{fig:13}
\end{figure*}

The calculation results are summarized in columns 9--11 of Table~\ref{table:1}. 
Low-velocity outflow, which is typically seen in simulations of supercritical clouds \citep[e.g.][]{tomisaka02,machida08,bate14,tomida15}, appears only in the magnetically supercritical models R2, R3 and R5. 
In the magnetically subcritical models, since the first core does not have sufficient rotation, low-velocity outflow does not appear. 
Although a high-velocity jet appears in all models, the momentum of the jet in the magnetically subcritical models is smaller than that in magnetically supercritical models (Fig.~\ref{fig:13}). 
Thus, the jet in magnetically subcritical models might be a transient phenomenon.

A rotationally supported disc appears in the magnetically supercritical models, which is consistent with recent works \citep{tsukamoto15,tsukamoto15b,tomida15,machida14a,machida16}. 
A very small rotating disc with a size of $<0.1$\,AU appears around the protostar in the magnetically subcritical models R03 and R05, while no rotating disc appears in model R01. 
A longer time integration is necessary to determine whether or not the disc and jet grow during the main accretion stage in the magnetically subcritical models.

\section{Discussion} 
\label{sec:discussion}
This study compares the star formation processes in magnetically supercritical and subcritical clouds under the same conditions by adopting the same settings for both types of cloud. 
We use common settings, typically adopted in this type of simulation (especially for simulations of collapsing magnetically supercritical clouds).
The calculation results show that the cloud collapse and star formation processes in magnetically subcritical clouds differ considerably from those for magnetically supercritical clouds. 
Comparing observations in star forming regions, our findings may be able to provide limits on the initial conditions of star formation and unveil the star formation process. 
However, there are some caveats which likely affect the results. 
In this section, we discuss the effects of the surrounding environment and other non-ideal MHD effects (especially the Hall effect) on the star formation process in magnetically subcritical clouds.

\subsection{Environmental Effects}
We assumed that the interstellar medium, which encloses the star forming cloud, has a low density, a uniform magnetic field and no rotation,  as described in \S\ref{sec:init}. 
The physical quantities for star forming cloud cores and their environment adopted in this study are in rough agreement with  observations. 
However, the assumption of a non-rotating interstellar medium might affect the results, especially for magnetically subcritical models.

The star forming cloud is connected to the interstellar medium through magnetic field lines. 
Thus, a part of the angular momentum in the star forming cloud is transferred to the interstellar medium by magnetic braking \citep[e.g.][]{mestel79,nakano90,basu94,basu95a,basu95b}. 
In magnetically supercritical clouds, however, cloud collapse and star formation should occur over a short duration $\sim t_{\rm ff}$ after the star forming cloud  forms \citep{inoue12}, where $t_{\rm ff}$ is the freefall timescale for the initial cloud. 
Thus, in this case, since there is not enough time to transfer the angular momentum from the star forming cloud to the interstellar medium, the angular momentum in the cloud is roughly conserved.  
Therefore, in such a cloud, the angular momentum transfer at the cloud scale does not greatly affect the cloud collapse and subsequent star formation processes \citep[][]{basu95b,machida11b}.

On the other hand, in magnetically subcritical clouds, it takes a long time to initiate a rapid collapse because cloud collapse does not begin until the magnetic flux around the cloud centre is sufficiently removed by ambipolar diffusion. 
Thus, there is sufficient time to transfer the angular momentum of the cloud to the interstellar medium by magnetic braking 
and, therefore, the angular momentum (or angular velocity) gradually decreases and approaches zero as shown in Figure~\ref{fig:3}. 
Our results are in good agreement with the pioneering works of \citet{mouschovias79,mouschovias80}, in which they adopted a non-rotating environment and showed a decreasing angular momentum in magnetically subcritical clouds.
However, when the interstellar medium has an angular momentum (or angular velocity), the angular velocity of the cloud approaches that of the interstellar medium. 
This is because a lump of gas inside the cloud corotates with the lumps of gas outside the cloud after a long time has passed since they were connected through the magnetic field lines \citep[e.g.][]{mestel79}. 
Thus, the evolution of a magnetically subcritical cloud might depend on the rotation rate of the background interstellar medium.

\begin{figure*}
\includegraphics[width=150mm]{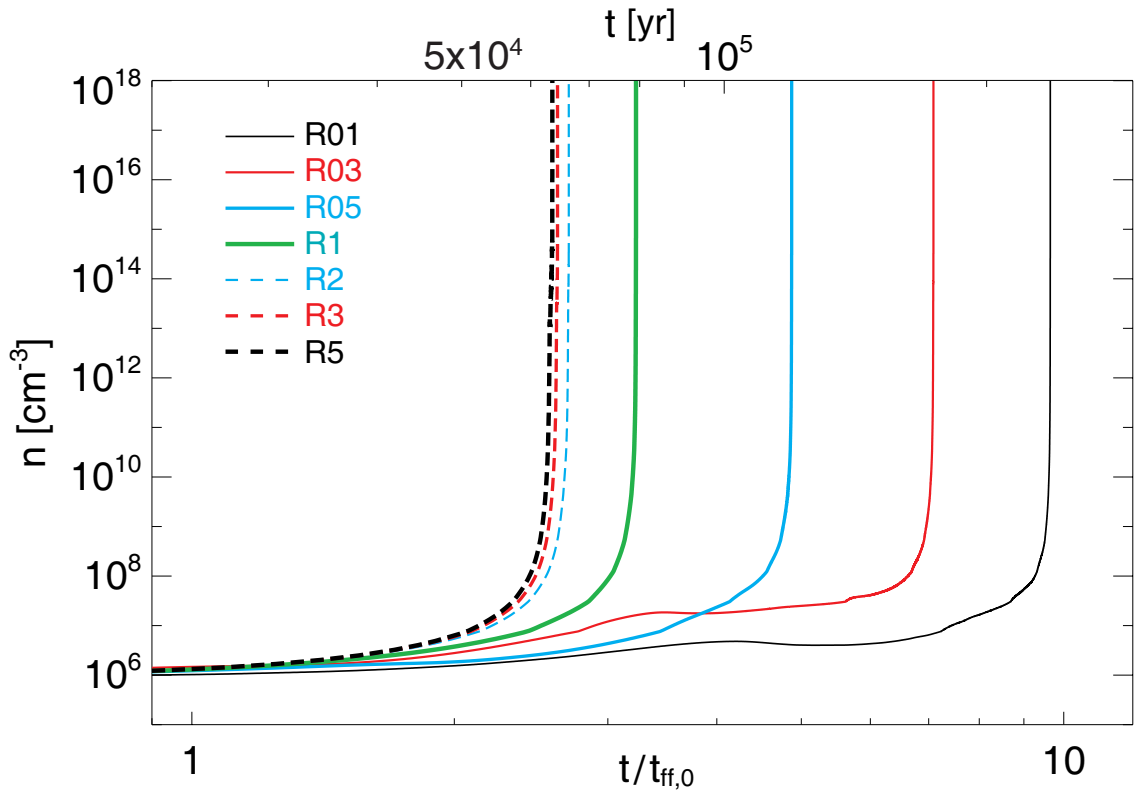}
\caption{
Central number density for each model plotted against the elapsed time normalized by the initial free fall timescale $t_{\rm ff,0}$ (lower axis) and that  in unit of years (upper axis). 
}
\label{fig:2}
\end{figure*}

\citet{basu94,basu95a,basu95b} included a background angular velocity of $\Omega_{\rm b}\sim10^{-15}$\,s$^{-1}$, which is derived from galactic rotation and is comparable to or slightly larger than the rotational rate of the magnetically subcritical models of this study just before the rapid collapse begins (Fig.~\ref{fig:2}). 
Although \citet{basu94,basu95a,basu95b} investigated only the early phase of star formation until the central density reaches $n_{\rm c}\sim 10^6\cm$, their results do not contradict ours.
Thus, unless the interstellar medium has an extremely large rotation rate, the results would not be changed when rotation of the interstellar medium is assumed.

In addition, \citet{basu95a} also investigated the effect of the interstellar density on magnetic braking and showed that the difference in density between a star forming cloud and the interstellar medium does not significantly affect the results. 
Thus, previous works indicate that environmental effects will not significantly change the results. 
However, to confirm previous works, we will investigate the evolution of a magnetically subcritical cloud with different environmental parameters in a forthcoming paper.

\subsection{Non-ideal MHD Effects}
 In this study, we considered Ohmic dissipation and ambipolar diffusion as non-ideal MHD terms. 
These effects dissipate or weaken the magnetic field. 
On the other hand, for non-ideal MHD effects, we ignored the Hall effect, which does not dissipate the magnetic field, but changes the direction of the magnetic field vectors. 
Recent studies showed that the Hall effect can redistribute the angular momentum in a low density region or around the first core \citep{tsukamoto15}. 
Our calculation showed that the lifetime of the first core is short due to the small angular momentum (\S\ref{sec:beforeps}).
When the Hall effect is considered, the rotation rate at small scales might be changed. 
However, although the Hall effect may change the angular momentum distribution, it cannot increase the total angular momentum in a collapsing cloud. 
Thus, the Hall effect might not significantly affect the evolution of magnetically subcritical clouds that have an extremely small total angular momentum.

In this study, we adopted a dust particle size of 0.1$\mu$m (\S\ref{sec:model}), while the coefficients of the non-ideal MHD terms (Ohmic, ambipolar and Hall coefficients) significantly depend on the dust properties. 
Thus, more realistically, we need to change or parameterize dust properties (size distribution, total mass, etc.) to check the results of this study. 
However, such an investigation is beyond the scope of the present study,
which compares the evolution of magnetically supercritical and subcritical clouds using the same dust model. 
We expect that changing the dust properties will not qualitatively change the results because the differences in the cloud evolution shown in this study are caused by largescale magnetic diffusion and magnetic braking, which do not quantitatively depend on dust properties \citep{basu94,basu95a,tassis07b}.
However, including  a Hall term with a more flexible dust model should allow us to investigate cloud evolution to more quantitatively determine the star formation process in magnetically supercritical and subcritical clouds, and such a study will be undertaken in a future work.

\subsection{Effects of Cloud Turbulence}
As described in \S\ref{sec:init}, we adopted simple initial prestellar clouds as the initial state to investigate the effect of magnetic field strength on cloud collapse and protostar formation. 
In the prestellar clouds, we set a coherent magnetic field and rigid rotation, and did not consider turbulence. 
The calculation results indicated that no disc appears in magnetically subcritical clouds, as shown in \S\ref{sec:results}.

On the other hand, \citet{seifried12,seifried13} showed that cloud turbulence promotes disc formation because the disordered magnetic structure and turbulent motion in the surroundings of the disc help alleviate the magnetic braking. 
Their picture is different from the classical picture of magnetic braking \citep[e.g.][]{mouschovias80}, in which an ordered magnetic field was assumed. 
Although \citet{seifried12,seifried13} only investigated the evolution of magnetically supercritical clouds with $\mu \ge 2.6$, a rotationally supported disc may form in magnetically subcritical clouds when sufficiently strong turbulence exists.
However, we cannot know whether turbulence is maintained in magnetically subcritical clouds until the magnetic flux is significantly removed by ambipolar diffusion. 
Besides, \citet{kudoh11} showed that the ambipolar diffusion timescale shortens when supersonic flow exists in magnetically subcritical clouds, because the cloud density is transiently enhanced and ambipolar diffusion becomes more effective in such a region. 

In this study, to focus on the effect of magnetic field strength on the cloud evolution, we adopted idealized prestellar clouds. 
In future studies, we will need to include turbulence in the calculation to more realistically investigate the evolution of magnetically subcritical clouds.

\section{Summary}
\label{sec:summary}
\begin{figure*}
\includegraphics[width=150mm]{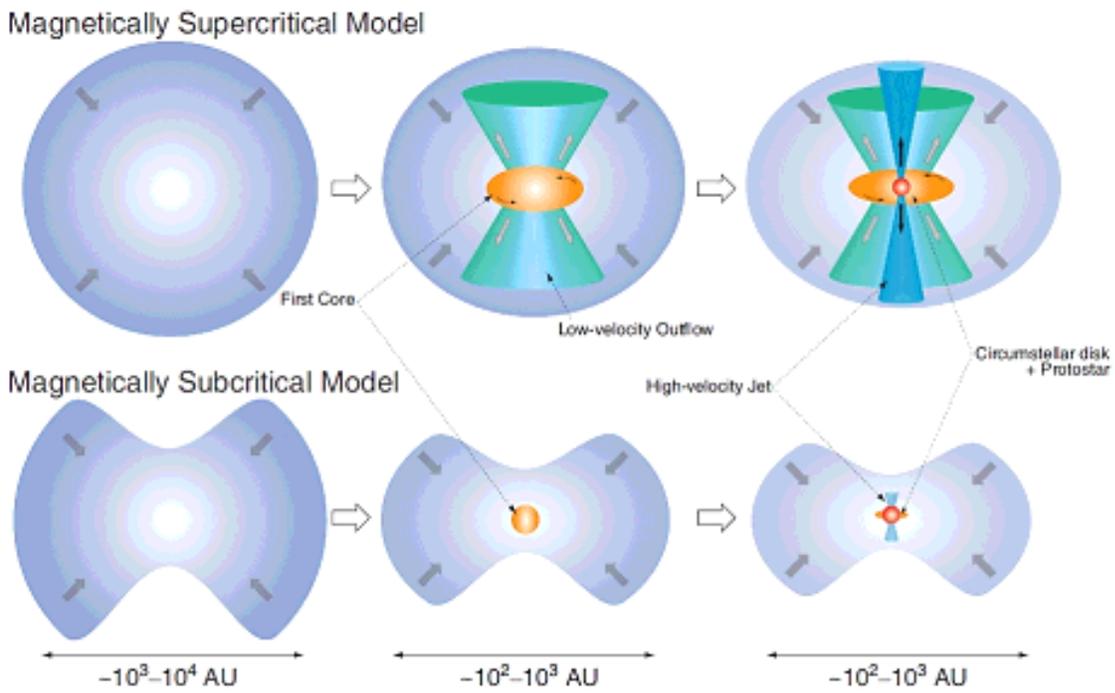}
\caption{
Schematic view of cloud evolution and star formation in magnetically supercritical (top) and subcritical (bottom) clouds.
}
\label{fig:14}
\end{figure*}

To investigate the star formation process in magnetically subcritical clouds and the differences in the cloud evolution between magnetically supercritical and subcritical clouds, we investigated the evolution of both types of cloud. 
Figure~\ref{fig:14} shows a schematic view of cloud evolution until just after protostar formation. 
As seen in the top panel, in magnetically supercritical clouds, a rotating first core forms and drives low-velocity outflow before protostar formation. 
Then, the angular momentum of the first core is transferred by magnetic effects such as low-velocity outflow and magnetic braking, while the magnetic field dissipates by Ohmic dissipation and ambipolar diffusion inside the first core (or a magnetically inactive region with an extremely low ionisation degree). 
Therefore, the Lorentz and centrifugal forces weaken, and a second collapse is induced due to the dissociation of molecular hydrogen after the excess angular momentum and magnetic flux are removed from the central region. 
After the central density reaches $n_{\rm c} \gtrsim 10^{20}\cm$ and the dissociation of molecular hydrogens is complete, the cloud collapse stops and a protostar is born.  
Around the protostar, since the gas temperature and ionisation degree are high, the magnetic field is again well coupled with neutral gas. 
Thus, magnetic field lines are strongly twisted due to the rotation of the protostar and disc inner edge, and a high-velocity jet appears.  
Several years after protostar formation, the first core (remnant) becomes a rotationally supported disc.  
The circumstellar disc mass is comparable to the protostellar mass, because the first core remnant is more massive than the protostar during the early mass accretion stage. 
Thus, gravitational instability is induced in the circumstellar disc. 
Both gravitational instability and magnetically driven jets contribute to the angular momentum transfer around the protostar, and the circumstellar disc gradually becomes stable. 
This picture of star formation in the magnetically supercritical clouds is now well established and widely accepted as a typical star formation scenario \citep[e.g.][]{inutsuka12}.

We next consider the star formation process in magnetically subcritical clouds obtained in this study (Fig.~\ref{fig:14} bottom panel). 
In these clouds, since the initial prestellar cloud is supported by the Lorentz force, a rapid gravitational contraction does not occur until the magnetic flux around the cloud centre is sufficiently removed by ambipolar diffusion. 
Before the rapid contraction, the whole cloud takes on a butterfly-like shape and a large fraction of the angular momentum in the cloud is transferred to the interstellar medium by magnetic braking during this epoch. 
After sufficient magnetic flux is lost, the cloud centre begins a rapid collapse and the first core forms, in the same manner as for the magnetically supercritical case. 
However, since a large fraction of the angular momentum is already removed from the collapsing cloud, the first core is mainly supported by the pressure gradient force and has a nearly spherical shape. 
As a result, the first core has a short lifetime due to the lack of rotation, and it cannot drive a low-velocity wide-angle outflow. 
In addition, because the first core does not acquire a sufficient angular momentum, a second collapse occurs a short time after the first core formation in the magnetically subcritical case.
A protostar forms following the second collapse and the first core remnant disappears several years after the protostar formation. 
Thus, no massive circumstellar disc evolves from the first core (remnant)  in the magnetically subcritical case.
On the other hand, weak jets appear around the protostar because the cloud contracts sufficiently and the rotational motion manages to catch up with the infall motion at very small scales due to the angular momentum conservation. 
However, in magnetically subcritical clouds, we cannot know whether the jet driving lasts for a long time. 
In addition, the formation and evolution of the circumstellar disc in magnetically subcritical clouds are not clear and cannot be revealed from this study.  
A long-term calculation of the main accretion phase is necessary to establish the star formation process in magnetically subcritical clouds.

This study has shown that the formation process of protostars in magnetically subcritical clouds differs considerably from that in magnetically supercritical clouds. 
Recent ALMA observations have shown that a large rotationally supported disc already exists around Class 0 protostars \citep{murillo13,sakai14,ohashi14,codella14,lee14,aso15}, while in some cases observers could not find a rotationally supported disc around protostars \citep{yen15,yen15b,yen17}. 
Although the lack of detection of a disc may be due to the limited spatial resolution of ALMA, a possible explain may lie in the way the protostellar system is formed in magnetically subcritical clouds. 
Thus, there may be different modes of star formation, as shown in Figure~\ref{fig:14}. 
Further theoretical and observational investigations are necessary to establish the star formation scenarios.

\section*{Acknowledgements}
We have benefited greatly from discussions with ~K. Tomisaka, ~K. Tomida, ~T. Nakano and ~S. Basu. 
We also thank the reviewer for many useful comments on this paper.
This work was supported by JSPS KAKENHI Grant Numbers JP25400232, JP15K05032 and JP17K05387, JP17H06360.
This research used computational resources from the high-performance computing infrastructure (HPCI) system provided by the Cyberscience Center, Tohoku University, and the Cybermedia Center, Osaka University, and the Earth simulator, JAMSTEC through the HPCI System Research Project (Project ID: hp150092,hp160079,hp170047).
Simulations were also performed by 2017 Koubo Kadai on Earth Simulator (NEC SX-ACE) at JAMSTEC.

\clearpage
\begin{table*}
\small{
\setlength{\tabcolsep}{3.5pt}
\begin{center}
\begin{tabular}{c|cccccccccc|ccccccc} \hline
{\footnotesize Model} & $M_{\rm cl}$ & $R_{\rm cl}$ & $B_0$ & $\Omega_0$ & \multirow{2}{*}{$\alpha_0$} & \multirow{2}{*}{$\beta_0$} & \multirow{2}{*}{$\mu$} 
& \multirow{2}{*}{Outflow} & \multirow{2}{*}{Jet}  &  \multirow{2}{*}{RSD}  \\
& {\scriptsize [$\msun$]} & {\scriptsize [AU]} & [$\mu$\,G] & {\scriptsize [$10^{-13}$\,  s$^{-1}$]} &  &  &  \\
\hline
R01 & \multirow{7}{*}{1.1} & \multirow{7}{*}{$6.2 \times 10^3$} &  1380 & \multirow{7}{*}{2.2}  & \multirow{7}{*}{0.47} & \multirow{7}{*}{0.01} & 0.1  & $\times$ & $\bigtriangleup$  & $\times$ \\
R03 &  &  & 436 &  &  &   & 0.3 & $\times$   & $\bigtriangleup$  & $\bigtriangleup$ \\
R05 &  &  & 261 &  &  &   & 0.5 & $ \times$  & $\bigtriangleup$  & $\bigtriangleup$ \\
R1 &   &  & 130 &  &  &   & 1   & $\times$   & $ \bigcirc$        & $\bigtriangleup$ \\
R2 &   &  & 65 &  &  &   & 2    & $\bigcirc$  & $ \bigcirc$        & $\bigcirc$ \\
R3 &   &  & 44 &  &  &   & 3    & $\bigcirc$  & $ \bigcirc$        & $\bigcirc$ \\
R5 &   &  &26  &  &  &   & 5    & $\bigcirc$  & $ \bigcirc$        & $\bigcirc$ \\
\hline
\end{tabular}
\end{center}
\caption{
Model parameters.
Column 1 gives the model name. 
Columns 2--5 give the cloud mass $M_{\rm cl}$, cloud radius $R_{\rm cl}$, magnetic field strength $B_0$ and angular velocity $\Omega_0$. 
Columns 6 and 7 give the ratios $\alpha_0$ and $\beta_0$ of the thermal and rotational energies to the gravitational energy of the initial cloud. 
Column 8 gives the initial mass-to-flux ratio $\mu$ normalized by the critical value.
Columns 9--11 describe whether a low-velocity outflow (Outflow), high-velocity jet (Jet)  and  rotationally supported disc (RSD) appear ($\bigcirc$) or not ($\times$), in which triangle ($\bigtriangleup$) indicates the appearance of a very small ($< 0.1$\,AU) disc or a very weak jet.
}
\label{table:1}
}
\end{table*}

\clearpage
\begin{figure*}
\includegraphics[width=150mm]{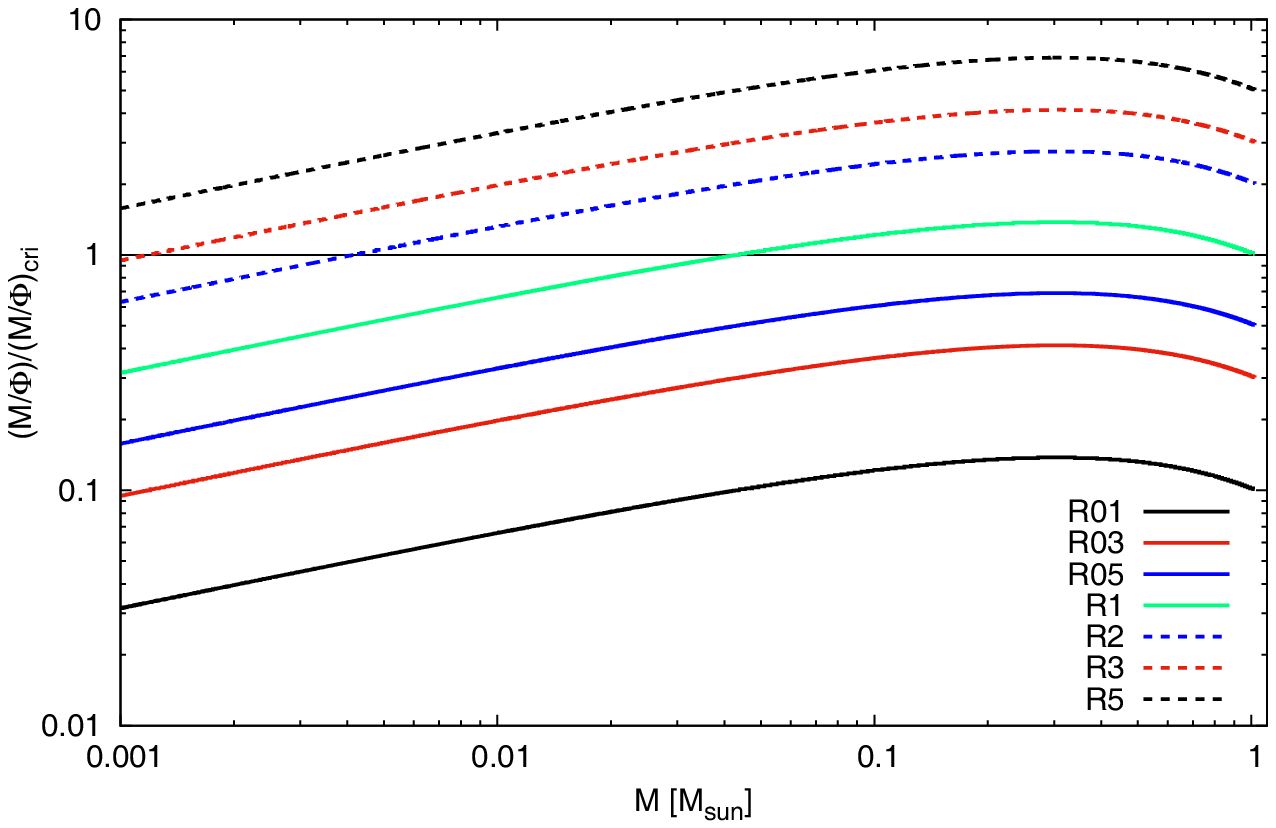}\\
\includegraphics[width=150mm]{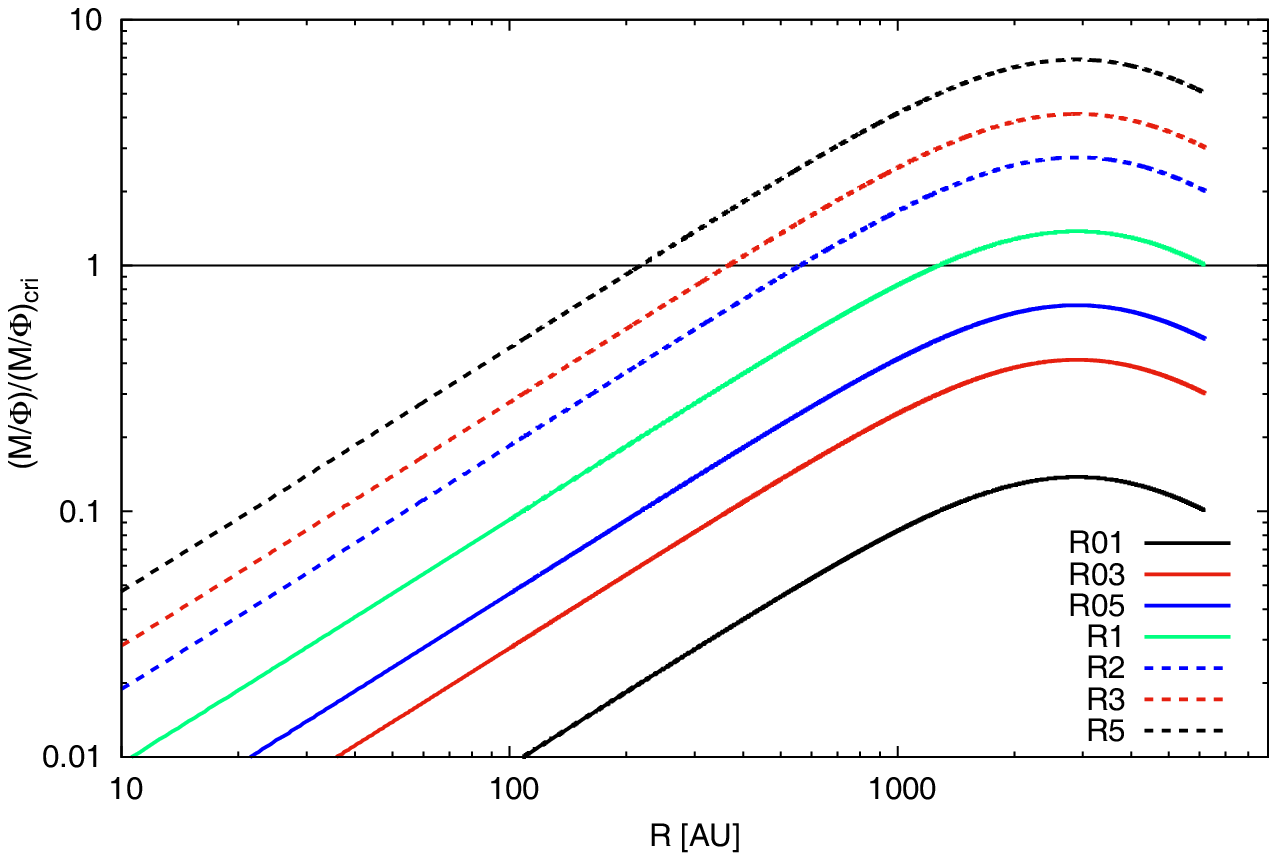}
\caption{
Initial mass-to-flux ratio for each model plotted against cumulative mass (top) and cloud radius (bottom).  
}
\label{fig:1}
\end{figure*}
\clearpage

\clearpage

\clearpage

\clearpage

\clearpage

\begin{figure*}
\includegraphics[width=150mm]{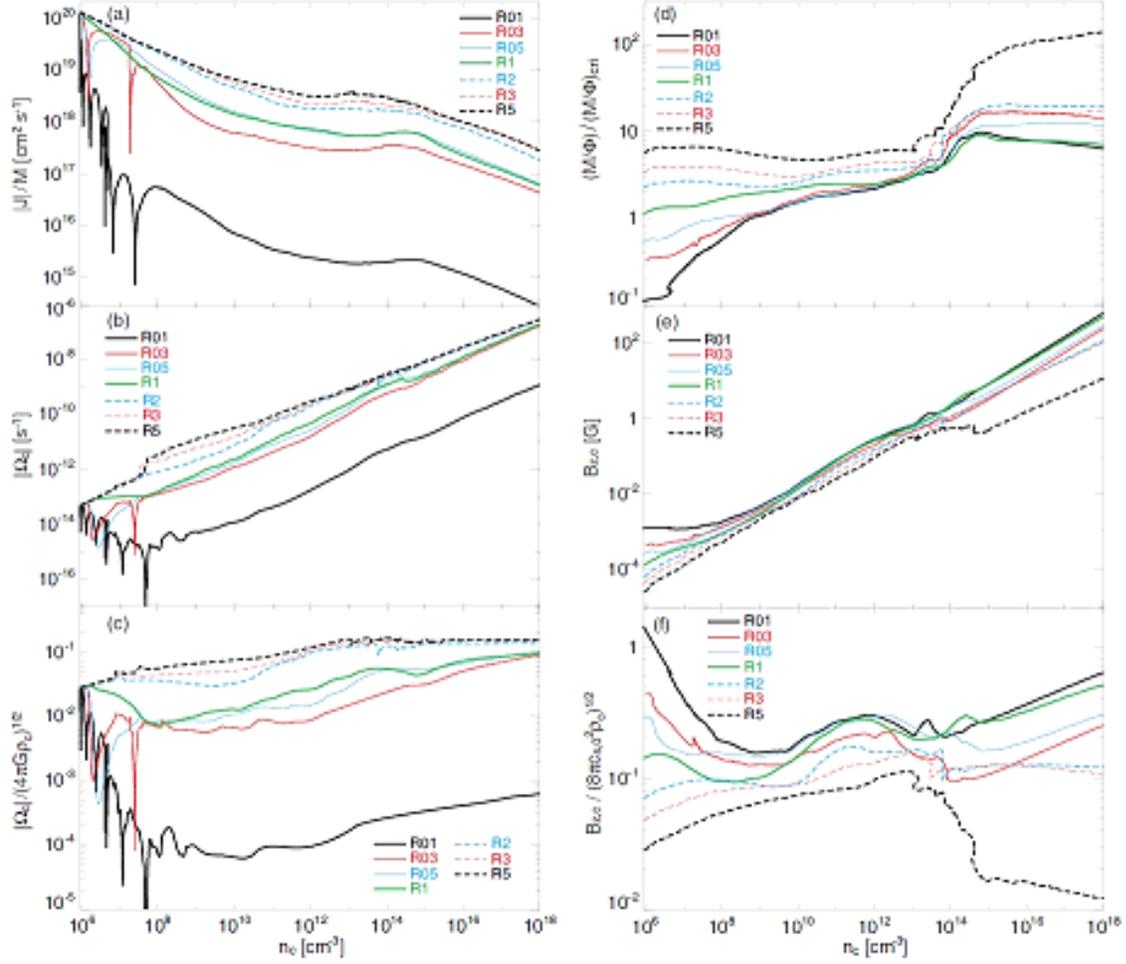}
\caption{
(a) Absolute value of the specific angular momentum in the region $\rho > 0.1 \rho_{\rm max}$. (b) Absolute value of the angular velocity at the cloud centre. (c) Absolute value of the angular velocity normalized by $(4\pi G \rho_{\rm c})^{-1/2}$. (d) Mass-to-flux ratio normalized by the critical value in the region of $\rho > 0.1 \rho_{\rm max}$. (e) Magnetic field strength at the centre of the cloud. (f) Magnetic field strength normalized by $(8\pi c_{s,0}^2\rho_{\rm c})^{1/2}$  at the centre of the cloud for all models.
}
\label{fig:6}
\end{figure*}
\clearpage

\clearpage

\clearpage

\clearpage


\clearpage

\clearpage

\clearpage

\clearpage

\appendix
\section{Heating by Magnetic Dissipation}
\label{sec:appendix}
\begin{figure*}
\includegraphics[width=150mm]{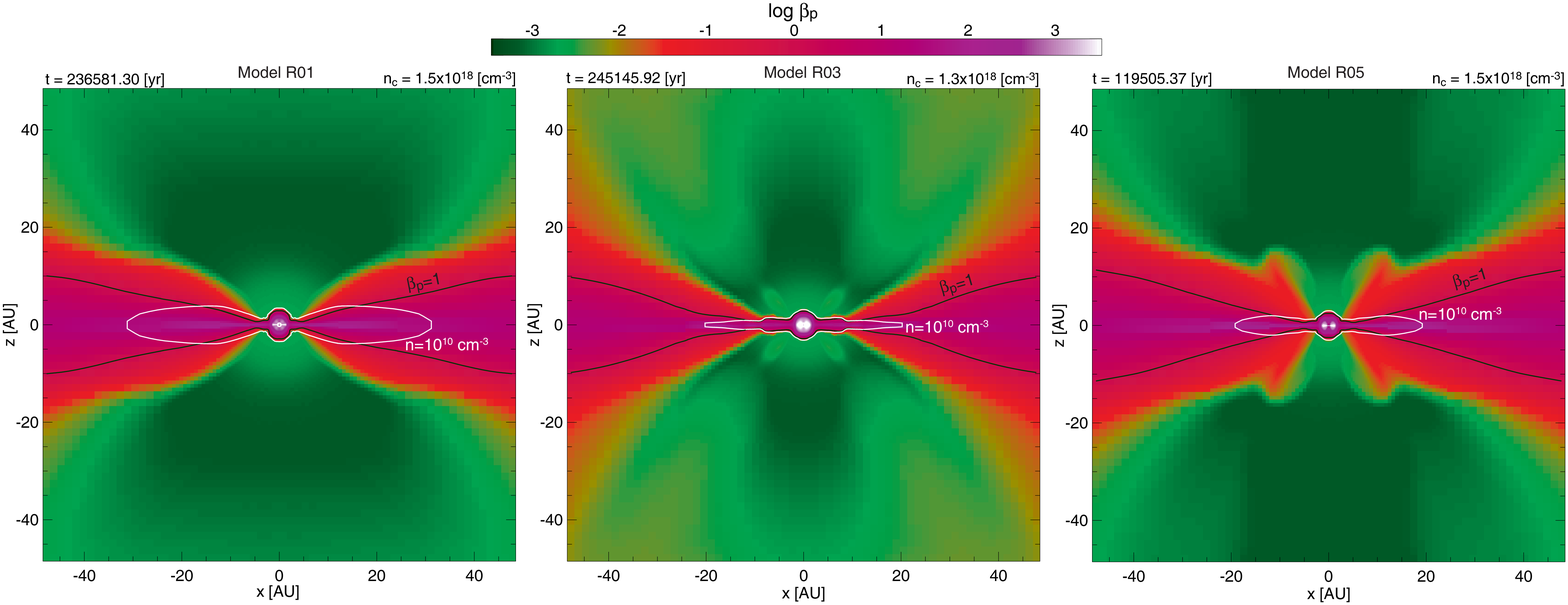}
\caption{
Plasma beta ($\beta_{\rm p}$) on the $y=0$ plane at the protostar formation epoch for models R01 (left), R03 (middle) and R05 (right). 
In each panel, white and black contours correspond to $n=10^{10}\cm$ and  $\beta_{\rm p}=1$, respectively.
}
\label{fig:A1}
\end{figure*}

This study ignored heating by ambipolar diffusion and Ohmic dissipation, because a simple barotropic equation of state was used, as described in \S\ref{sec:model}. 
For magnetically subcritical models, heating by ambipolar diffusion in the low density region ($n\ll 10^{10}\cm$) should be ignored because the thermal cooling due to dust grains is very efficient.
On the other hand, heating by Ohmic dissipation and ambipolar diffusion may influence the dynamical evolution in the high-density region of $n\gtrsim 10^{10}\cm$. 
To confirm this, the plasma beta $\beta_{\rm p}$, which is defined as
\begin{equation}
\beta_{\rm p} = \dfrac{8\pi c_s^2 \rho}{B^2},
\end{equation}
for subcritical models R01, R03 and R05 is plotted in Figure~\ref{fig:A1}. 
Both ambipolar diffusion and Ohmic dissipation become effective when the number density exceeds $n\gtrsim 10^{10}\cm$ \citep{nakano02}, which corresponds to the region inside the white contour in Figure~\ref{fig:A1}.
The figure indicates that the dissipation of magnetic field mainly occurs in the region of $\beta_{\rm p}\gtrsim 1$ where the thermal energy  dominates the magnetic energy.
Thus, it is expected that the dynamical structure in the magnetically inactive region is not significantly changed when the heating by ambipolar diffusion and Ohmic dissipation is taken into account.

However, the heating by ambipolar diffusion and Ohmic dissipation should increase the lifetime of the first core because the first core is mainly supported by the thermal pressure gradient force against the self-gravity \citep{larson69,masunaga00}.
Note that the first core is partly supported by the centrifugal force in magnetically supercritical models \citep{saigo06,saigo08}. 
Since the dissipation of the magnetic field mainly occurs inside the first core, the magnetic field might become weaker in calculations that consider the heating by magnetic dissipation than in those that consider a barotropic equation of state due to the long lifetime of the first core. 
The dissipation of magnetic field is effective as long as the first core exists in the gas collapsing phase.

\section{Rate of Ambipolar diffusion and Ohmic dissipation}
\label{sec:A2}
This study investigated the evolution of both magnetically supercritical and subcritical clouds, in which ambipolar diffusion and Ohmic dissipation play an important role.  
The coefficients of ambipolar diffusion ($\eta_{\rm A}$ in eq.[\ref{eq:reg}]) and Ohmic dissipation ($\eta_{\rm O}$ in eq.[\ref{eq:reg}]) are considered to depend strongly on dust properties, chemical abundance and ionisation sources, which are not well determined by both theoretical and observational studies at present.  
Thus, we did not include details about them. 
Since our results may be quantitatively changed when we adopt different settings (e.g. different dust properties) to calculate $\eta_{\rm A}$ and $\eta_{\rm O}$, we show only the qualitative results in this paper.

In this section, we briefly describe our settings for the calculation of $\eta_{\rm A}$ and $\eta_{\rm O}$, and then compare them.
The calculation was done according to \citet{umebayashi80}, \citet{nakano86}, \citet{nishi91}, \citet{nakano02} and \citet{okuzumi09}.
We considered the chemical species e, H$_3^+$, HCO$^+$, Mg$^+$, He$^+$, C$^+$ and H$^+$, and considered the dust grains to be changed particles. We also included thermal ionisation of potassium. 
The abundance of each species and chemical network are almost the same as that used in \citet{umebayashi80}.
A dust-to-gas ratio of 0.016 was adopted with a grain size of $0.1\,\mu$m.
We adopted a cosmic-ray ionisation rate of $\zeta = 1.0\times10^{-17}$s$^{-1}$ \citep[for details, see][]{tomida15}.
The equations for calculating $\eta_{\rm A}$ and $\eta_{\rm O}$ can be found in \citet{umebayashi80}, \citet{nakano86}, \citet{umebayashi90}, \citet{nishi91} and \citet{nakano02}.

\begin{figure*}
\includegraphics[width=150mm]{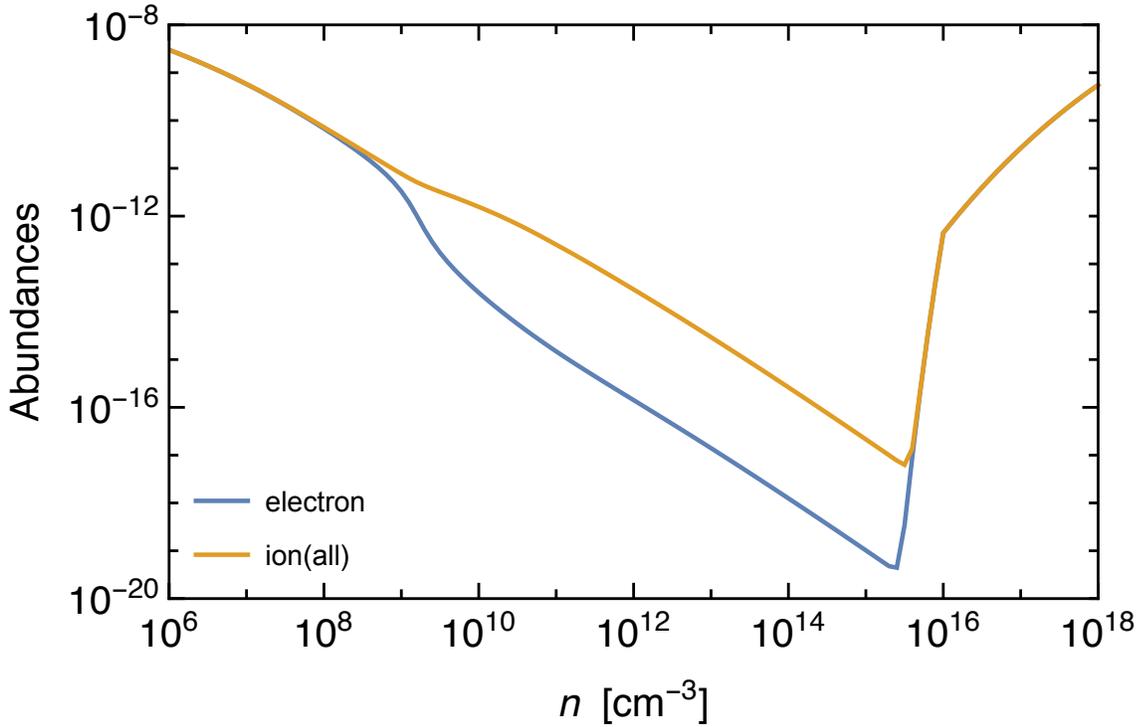}
\caption{
Abundances of electrons and ions against the number density.
All the charged species are summed in the `ion' value.
}
\label{fig:A2}
\end{figure*}

\begin{figure*}
\includegraphics[width=150mm]{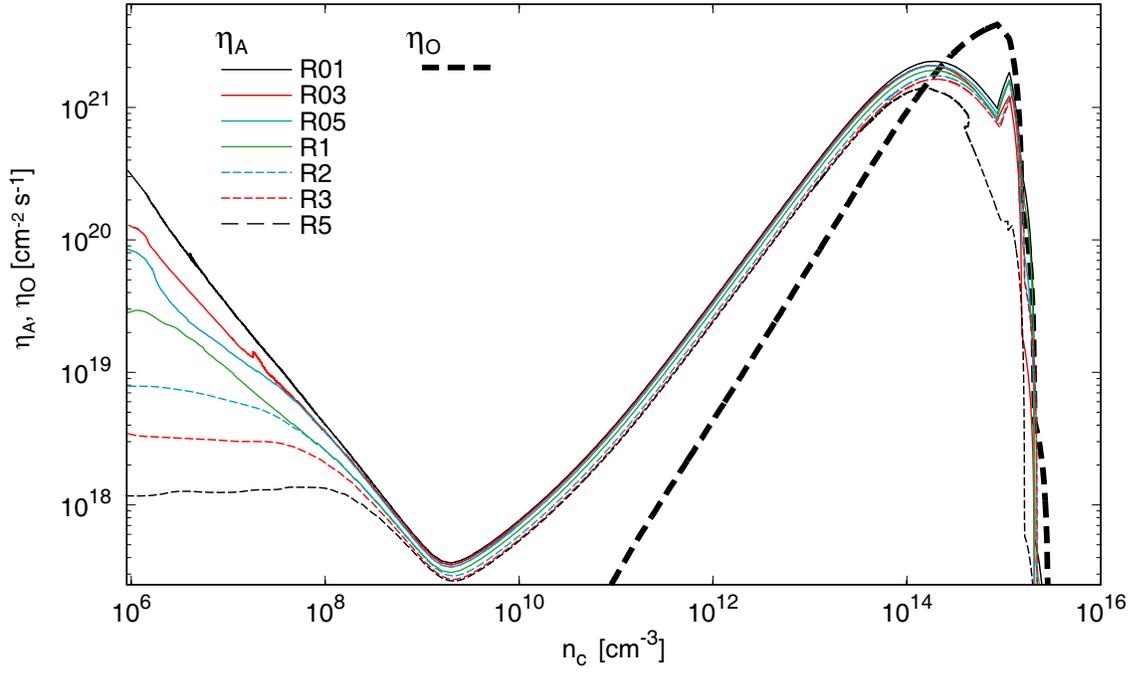}
\caption{
Coefficients of ambipolar diffusion $\eta_{\rm A}$ for all models and Ohmic dissipation $\eta_{\rm O}$ against the number density.
}
\label{fig:A3}
\end{figure*}

Figure~\ref{fig:A2} shows the abundance of changed ions and electrons, in which the ion abundance is the sum of all species (for details, see \citealt{umebayashi90}).  
A sharp increase in both abundances at $n_c \sim 10^{15}\cm$ is due to the thermal ionisation of potassium.
Figure~\ref{fig:A3} shows $\eta_{\rm A}$ and $\eta_{\rm O}$ used in the non-ideal MHD calculations. 
Since the same barotropic equation of state was used in all models (\S\ref{sec:model}) and Ohmic resistivity does not depend on the magnetic field strength, $\eta_{\rm O}$ is the same among the models.
On the other hand, the coefficient of ambipolar diffusion $\eta_{\rm A}$ depends on the magnetic field strength.
Thus, a noticeable difference can be seen in the low-density region of $n_c \lesssim 10^8\cm$ among the models. 
However, there is a slight difference in $\eta_{\rm A}$ in the region $n_c \gtrsim 10^{10}\cm$, because the magnetic field strengths in this range do not significantly differ among the models, as seen in Figure~\ref{fig:6}{\it d}. 
The figure indicates that Ohmic dissipation dominates ambipolar diffusion in the range $\gtrsim 10^{14}\cm$.
Thus, both ambipolar diffusion and Ohmic dissipation are important to correctly investigate the evolution of the magnetic field in collapsing clouds.

\end{document}